\documentclass[a4paper,11pt]{article}
\def\prd{Phys. Rev. D}

\def\apj{Astrophys. J.}

\def\jcap{JCAP}

\pdfoutput=1
\usepackage{jcappub,graphicx,epsfig,natbib,color,times,bm,amsmath,multirow}
\usepackage[ddmmyyyy,hhmmss]{datetime}

\begin{document}
\title{A blind search for a common signal in gravitational wave detectors}

\author[a,b]{Hao Liu,}\emailAdd{liuhao@nbi.dk}

\author[a]{James {Creswell},}\emailAdd{dgz764@alumni.ku.dk}

\author[a]{Sebastian {von Hausegger},}\emailAdd{s.vonhausegger@nbi.dk}

\author[c]{Andrew {D. Jackson}}\emailAdd{jackson@nbi.dk}

\author[a]{and Pavel Naselsky}\emailAdd{naselsky@nbi.dk}

\affiliation[a]{The Niels Bohr Institute \& Discovery Center, Blegdamsvej 17, DK-2100 Copenhagen, Denmark}

\affiliation[b]{Key laboratory of Particle and Astrophysics, Institute of High Energy Physics, CAS, 19B YuQuan Road, Beijing, China}

\affiliation[c]{The Niels Bohr International Academy, Blegdamsvej 17, DK-2100 Copenhagen, Denmark}

\abstract{
We propose a blind, template-free method for the extraction of a common signal
between the Hanford and Livingston detectors and apply it especially to the
GW150914 event. We construct a log-likelihood method that maximizes the
cross-correlation between each detector and the common signal and minimizes
the cross-correlation between the residuals. The reliability of this method is
tested using simulations with an injected common signal. Finally,
our method is used to assess the quality of theoretical gravitational wave templates for GW150914.} 


\maketitle

\section{Introduction}\label{sec:intro}
To date five gravitational wave (GW) events from binary black hole mergers
have been announced by the LIGO/Virgo collaboration~\cite{
PhysRevLett.116.061102, PhysRevLett.116.241103, PhysRevLett.118.221101,
PhysRevLett.119.141101, 2017arXiv171105578T}, among which GW150914 is the most
statistically significant. This event provides a unique opportunity for the
investigation of both templates and noise~\cite{2017JCAP...08..013C}. In spite
of the extensive discussion in the literature devoted to the exploitation of
the physical consequences of this observation, remarkably few papers have
addressed the morphological properties of the signal and the significance of
its detection~\cite{2016JCAP...10..014L,2016JCAP...08..029N}. Unlike the other
LIGO events (GW151226, GW170104, GW170814 and the most recent GW170817),
GW150914 can be identified in the time ordered data stream without the use of
the templates~\cite{2016JCAP...08..029N}. In this case, the use of templates
can be restricted to the determination of the masses, spins and distance to
the presumed black hole binary system. To the extent that assumptions
regarding the nature of the event are correct and the properties of the noise
are well understood, these physical parameters can be estimated using maximum
likelihood methods. However, searches for a common signal using methods that
are based solely on the use of templates have the potential danger of
misidentifying transients and/or systematic effects as part of the desired
signal. At the very least, preconceptions can introduce a bias in the
estimation of the inferred parameters. Such limitations on the reliability of
template-based signal extraction methods emphasize the desirability of
template-free methods.

In this paper we propose such a template-free method for determining the best
common signal in the Hanford and Livingston detectors. While nothing prevents
application of this method to any event with a sufficiently high
signal-to-noise ratio, in this paper we focus on the 0.2 s of the GW150914
event. The physical basis of the method is that there exists a common signal
in the LIGO Hanford and Livingston detectors. The presence of this signal,
$A(t)$, in each of the two detectors is characterised by the cross-correlation
coefficients $C_{AD}$, giving the residuals $R_D(t)=D(t)-C_{AD}\cdot A(t)$,
where $D(t)$ is the strain data for Hanford ($D=H$) or Livingston ($D=L$). Our
objective is to find $A(t)$ such that the cross-correlation between $A(t)$ and
each of the two data streams $D(t)$ is maximized, while the cross-correlation
between $R_H(t)$ and $R_L(t)$ is minimized. For this purpose, a maximum
likelihood approach for the cross-correlation functions is constructed by
means of their Fisher transforms~\cite{doi:10.1093/biomet/10.4.507}, in
combination with a random search for the optimal solution $A(t)$.

An important feature of our approach is that, unlike template fitting, we
obtain a family of optimal solutions, $A_i(t)$, that differ from one another
because of chance correlations between $A_i(t)$ and each of $R_{H,i}(t)$ and
$R_{L,i}(t)$, and between $R_{H,i}(t)$ and $R_{L,i}(t)$. These chance
correlations are an unavoidable part of the optimization technique, and they
reflect the intrinsic statistical properties of the problem.

Other template-free methods have been designed for detection and
identification of gravitational waves, such as the oLIB
method~\citep{PhysRevD.95.104046}, the X-pipeline
method~\citep{1367-2630-12-5-053034}, and the BayesWave
method~\citep{2015CQGra..32m5012C, 2017ApJ...839...15B}. Our method is
intended to be used as a follow-up to extract the common signal from detection
candidates. Unlike other approaches, it is completely blind and does not
explicitly depend on the amplitude.

Also note that in this work, we use the Hanford detector as the default
projection (see eq.~\ref{equ:template transit}), so all figures will be
suitable for comparison with Hanford data.

The outline of the paper is as follows. In section~\ref{sec:definitions and
likelihood} we present definitions and construct the likelihood function. In
section~\ref{sec:search} we introduce the necessary modifications of the data
streams to apply the method to LIGO data and describe the algorithm. The
non-uniqueness of the solution due to chance correlations is discussed in
section~\ref{sec:oscillation and solution} including first results, followed
by testing for the reliability of the method by the use of simulations in
section~\ref{sec:test by simulation}. We then investigate correlations between
the residuals in section~\ref{sec:resi-test}, and in
section~\ref{sec:Is_GW_good_for_common_signal} we evaluate binary black hole
merger templates for their goodness of fit to the common signal.

\section{Definitions and likelihood approach}\label{sec:definitions and likelihood}

\subsection{Cross-correlations and residuals}\label{sub:crosscor and res}

Assume two data sets $X(t)$ and $Y(t)$ (e.g. strain data from two independent
detectors) which contain a common signal $A(t)$. For convenience, we assume
that in the time range of interest $X$, $Y$, and $A$ have their average value
shifted to zero and their variance normalized to unity. With this
simplification the cross-covariance and the Pearson cross-correlation
coefficient of two vectors of size $N$ are equal.\footnote{Throughout this
paper we will change between the notations $X=X(t)$ and $X_i=X(t_i)$ at
convenience.}
\begin{equation}\label{equ:cc}
S_{XY}=\frac{1}{N-1}\sum_{i=1}^{N}{X_i\cdot Y_i} \qquad\quad C_{XY} =
\frac{S_{XY}}{\sqrt{S_{XX}S_{YY}}} = S_{XY}.
\end{equation}
We calculate the residuals according to linear regression of $A(t)$ against
$X(t)$ or $Y(t)$:
\begin{subequations}\label{equ:residual}
\begin{equation}
R_X = X - A\cdot\frac{S_{AX}}{S_{AA}} = X - A\cdot S_{AX},
\end{equation}
\begin{equation}
R_Y = Y - A\cdot\frac{S_{AY}}{S_{AA}} = Y - A\cdot S_{AY},
\end{equation}
\end{subequations}
where in the last steps we have again made use of the signals having been scaled to
unit variance. It is important to note that, by construction, the correlations
of both $R_X$ and $R_Y$ with $A$ are zero. Since it might be 
expected that the residuals are defined as the difference between the data and 
the best common signal, the factors of $S_{AX}$ and $S_{AY}$ in 
eqs.~\ref{equ:residual} may 
seem surprising.  This expectation can be fulfilled by a simple rescaling of $A$ 
which will have no influence on the results of the following procedure which is 
based solely on morphology. Note, too, that the amplitude of template (i.e., 
the analogue of the best common signal) is also freely adjustable in LIGO's 
analysis.

The criterion for our blind estimation of $A(t)$ now is to maximize $C_{AX}$
and $C_{AY}$ while simultaneously minimizing the cross-correlation between the
residuals, $C_{R_X R_Y}$. These can be obtained straightforwardly from
eqs.~\ref{equ:cc}. Note that in $C_{R_X R_Y}$ the residuals $R_X$ and $R_Y$
are not automatically normalized:
\begin{equation}\label{equ:cc_residual}
C_{R_X R_Y} = \frac{S_{R_X R_Y}}{\sqrt{S_{R_X R_X}\cdot S_{R_Y R_Y}}} \\
= \frac{S_{XY}-S_{AX}\cdot S_{AY}}{\sqrt{(1-S_{AX}^2) (1-S_{AY}^2)}}.
\end{equation}

\subsection{The likelihood approach}\label{sub:likelihood}

In order to make statements about the likelihood of the resulting correlations, we
need to estimate their distribution function. For simplicity, we obtain
approximate Gaussianity by using the Fisher
transformation~\citep{doi:10.1093/biomet/10.4.507}:
\begin{equation}\label{equ:fisher transform}
Z_{XY} = \frac{1}{2}\log\left(\frac{1+C_{XY}}{1-C_{XY}}\right),
\end{equation}
where $C$ is the Pearson cross-correlation coefficient from the previous
section.

We now define the log-likelihood function that we will use in the remainder of
this work:
\begin{equation}\label{equ:likelihood}
\log(L) = (Z_{AX}-E_{AX})^2+(Z_{AY}-E_{AY})^2 - k(Z_{R_X R_Y}-E_{R_X R_Y})^2,
\end{equation}
where $E$ is the expectation of $Z$ and $k$ is a weighting factor, introduced
to adjust the relative importance of the low correlation in the residuals over
those of the data sets $X$ and $Y$ with $A$. A larger value of $k$ places
greater emphasis on suppressing the residual correlation. Empirically, we find
that setting $k=8$ works well, i.e. the correlations of the residuals are low
and those of the common signal with the data are high, so we will keep this
choice throughout this work.\footnote{As the amplitude of the residuals can be
small this makes no statement about the rough shape of the common signal. We
offer a brief comparison of different choices of $k$ in section~\ref{sec:test
by simulation}.}

\sloppypar{As mentioned previously, we expect no correlation between $R_X$ and
$R_Y$, so we can set ${E_{R_XR_Y}=0}$ as the part of the initial null
hypothesis that we wish to accept. In the search for a common signal, $A$, we
similarly define the null hypothesis that $A$ is uncorrelated with each of
\mbox{$X$ and $Y$,} and therefore fix $E_{AX}= E_{AY}=0$. However, we wish to
reject this null hypothesis, which leads us choose the sign of the first two
terms in eq.~\ref{equ:likelihood} to be opposite to that of the third term.
Given these considerations, eq.~\ref{equ:likelihood} reduces to\footnote{It is
straightforward to generalize to three or more detectors by adding the
corresponding terms to this equation.}:}

\begin{equation}\label{equ:likelihood_simp}
\log(L) = Z_{AX}^2 + Z_{AY}^2 - k Z_{R_X R_Y}^2.
\end{equation}

A search based on eq.~\ref{equ:likelihood_simp} also allows for an unwanted,
peculiar solution
\begin{equation}\label{equ:unwanted solution}
A(t) = X(t)+O(t),
\end{equation}
where $|O(t)|\ll |X(t)|$ and is uncorrelated with $R_Y(t)$. In this case,
$Z_{AX}$ will be very large, while $Z_{R_XR_Y}$ will still be small, resulting
in an unreasonably large value of $\log(L)$. The common signal solution will
tend to be close to either Hanford or Livingston data, leaving almost
negligible residuals in the corresponding detector. This will result in an
inhomogeneity of residuals that is inconsistent with the properties of noise
observed outside the GW150914 domain. Therefore, for any solution $A$, with
$C_{AX}$ or $C_{AY}>0.9$,\footnote{Eq.~\ref{equ:likelihood} is a
standard definition of the log-likelihood function that requires 6 priors (3
expectations and corresponding RMS values). Although empirical priors are
widely used in likelihood analysis, following the discussions in
Sec.~\ref{sub:likelihood}, we have reduced the number of empirical priors from
6 to 2. One is $k$, and the other is the threshold, and both are tested to
ensure that the results are not very sensitive to them (see
section~\ref{sec:test by simulation} and section~\ref{sub:for the threshold}).
The threshold also ensures that the integrated probability/likelihood is
finite, which is required by self-consistency.} we reset $C_{AX}$ and/or
$C_{AY}$ to $0.9$ to reduce preference of this type of solution. Note that
this readjustment only affects the likelihood; it does not change the
estimation or the data. Thus the actual values for $C_{AX}$ and/or $C_{AY}$
can be higher than this limit, but the increase in correlation will not make
the associated common signal more preferable in the search.

It should be noted that, since we base our search on computations of
Pearson's cross-correlation coefficients, any amplitude difference between the
two signals $X(t)$ and $Y(t)$ (as could be the case for signals from two
gravitational wave detectors) is irrelevant. The search is expected to extract
the most probable common signal regardless of the observed relative amplitude.

The criterion presented here for defining the best common signal is not
unique. For example, the correlators $C_{A R_X}$ and $C_{A R_Y}$ are not
expected to be strictly zero as a consequence of chance correlations. It would
thus also be possible to replace the constraints imposed in
eqs.~\ref{equ:residual} with additional terms proportional to $Z^2_{A R_X}$
and $Z^2_{A R_Y}$ in eq.~\ref{equ:likelihood_simp}.  In situations where the
signal-to-noise ratio in $X$ and $Y$ differ significantly, it could be useful
to adjust the relative weight of the first two terms in
eqs.~\ref{equ:likelihood} and~\ref{equ:likelihood_simp} or to make independent
adjustments of the upper limits for $C_{AX}$ and $C_{AY}$.  Any of these
changes would introduce additional physically meaningful parameters in the
likelihood function.  Their presence would complicate the discussion but would
not alter the principle of the method proposed here.

\section{The search algorithm applied to Hanford and Livingston
data}\label{sec:search}

\subsection{Pre-processing of Hanford and Livingston data}\label{sub:pre-match}

In order to apply the ideas above to the strain data $H(t)$ and $L(t)$ from
the LIGO Hanford and Livingston detectors, respectively, it is necessary to
pre-process the data by band-passing the data to the frequency range of
interest\footnote{We frequency filter the data via a fourth-order (back-and-
forth) Butterworth filter which passes frequencies between 35 and 350~Hz and
additionally notch filter the narrow resonances, in accordance with
ref.~\cite{PhysRevLett.116.061102}.} and subsequently correcting for the phase
difference $\Delta$ and the time lag $\tau$.\footnote{See also the LIGO
tutorial at \url{http://losc.ligo.org/s/events/GW150914/LOSC_Event_tu
torial_GW150914.html}.} This can be done by modifying the Livingston Fourier
coefficients $L(\omega)$ into
\begin{equation}\label{equ:template transit}
\tilde{L}(\omega) = \alpha L(\omega) e^{i (\Delta + \omega \tau)},
\end{equation}
where $\Delta$ and $\tau$ are chosen to ``match'' $\tilde{L}(t)$ with $H(t)$.
The constant $\alpha$ represents the difference in the detector efficiency,
which is of no relevance in our method.

\begin{figure}[!htb]
 \centering
 \includegraphics[width=0.5\textwidth]{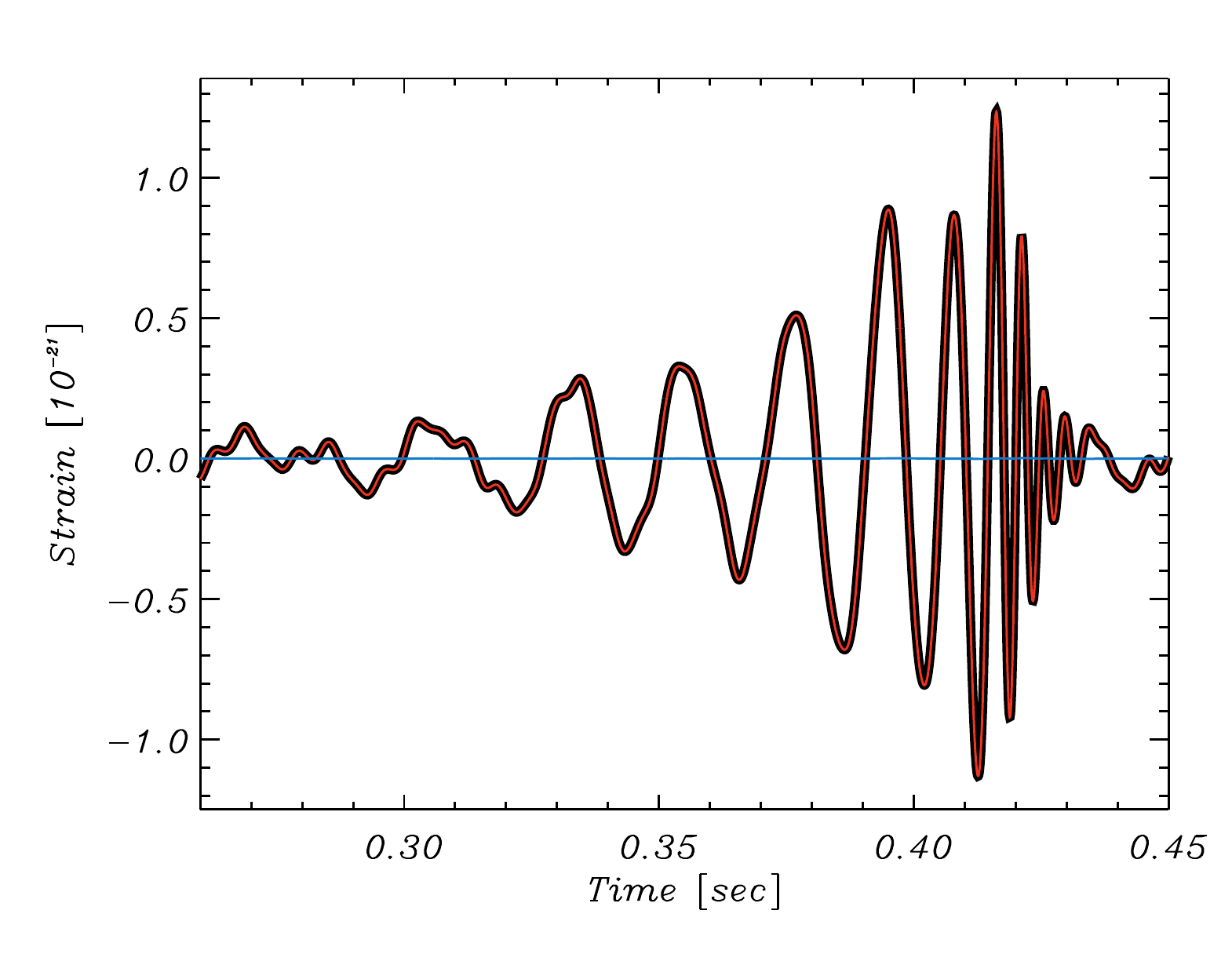}
 \caption{Matching the released Livingston template (red) to the Hanford
 template (black) by eq.~\ref{equ:template transit} with $\alpha=1.23$,
 $\tau=6.96$ ms, and $\Delta=2.72$. Their difference after matching is shown
 in blue. Their cross correlation before matching is about 0.41; after
 matching it is greater than 0.99. }
 \label{fig:ligo tpl conv}
\end{figure}

As an example, the LIGO Livingston template\footnote{Throughout this work we
will refer to this template as the ``GW150914 template". This is not LIGO's
best-fit template, however recent work \cite{Degeneracy} has shown that the
filtered templates are remarkably similar for a wide range of black hole
parameters. The effect of using templates based on different parameters is
studied in section \ref{sec:Is_GW_good_for_common_signal}.} presented in
figure~1 of~\citep{PhysRevLett.116.061102} can be converted to the Hanford
template by eq.~\ref{equ:template transit} with $\alpha=1.23$, $\tau=6.96$ ms,
and $\Delta=2.72$, as shown in figure~\ref{fig:ligo tpl conv}. However, by
matching the real Livingston data to Hanford's, we find $\tau=7.08$~ms and
$\Delta=2.86$, only marginally different than the values for the templates. In
this work, we will use $\tau=7.08$~ms and $\Delta=2.86$ for pre-processing;
the final result is insensitive to such small differences.

The matching described and illustrated above is central to much of LIGO/Virgo's
matched filter technique, including the primary results regarding GW150914
\cite{PhysRevD.85.122006,PhysRevD.93.122003}. However, the validity of the use
of matched filters requires several assumptions regarding the underlying
gravitational wave signal that are not true in general. In particular, the
gravitational waves emitted by a binary system that is precessing (i.e. with
component spins not aligned with the orbit axis) or a binary system with non-zero
orbital eccentricity will induce strains in two detectors that are not
equivalent after matching. Similarly, higher-order modes in the
gravitational waves will result in differences between the strain data in two
detectors that cannot be absorbed by the modifications described in
eq.~\ref{equ:template transit}. In the presence of these complications, a
common signal, $A(t)$, between the processed Hanford and Livingston strain
data only exists approximately.

Furthermore, it is possible that there exist ``foreground'' components that
introduce correlations in the Hanford and Livingston detectors that cannot be
captured by a single common signal. For a qualitative exposition see the
discussion in section~\ref{sec:resi-test}.

\subsection{The search algorithm}\label{sub:search process}

With the pre-processed data and the log-likelihood function presented in
eq.~\ref{equ:likelihood_simp}, it is possible to run a point-by-point search
for the best common signal. The algorithm proceeds iteratively as follows: we
start with an initial guess of $A(t)$ which is simply random white noise. One
after another, each data point is then moved by a small step $d(t)$ to produce
a new signal,
\begin{equation}\label{equ:one-point variance}
A'(t) = A(t)+ d(t)\cdot\delta(t,t_i),
\end{equation}
where $\delta(t,t_i)$ is a discrete Dirac delta function.

If $d(t)$ is sufficiently small, the change in the log-likelihood function
will be linear. Thus if the likelihood function increases, we accept $+d(t)$
as the movement direction, otherwise we move the point in the opposite
direction, $-d(t)$. This procedure is performed and recorded for each point
independently. However, all changes are applied only after all directions have
been determined, i.e. we ignore second- and higher-order partial derivatives
like $\partial^2 L/\partial A_i\partial A_j$ within one iteration. The updated
estimation of $A(t)$ is then used for the next iterative step.

Note that all analysis in this work can be done equivalently in either the
time or frequency domain, where we have better control of the degrees of
freedom (DOF). Since the data to be used is band-passed, we prefer to work in
the frequency domain. However, the final results will be presented in the time
domain for ease of understanding. Note, too, that at every iteration we
correct the signals to obey the zero mean and unit variance conditions
mentioned in section~\ref{sub:crosscor and res}.

\section{Oscillations in the likelihood function}\label{sec:oscillation and solution}

\begin{figure*}[!htbp]
  \centering
  \includegraphics[width=0.49\textwidth]{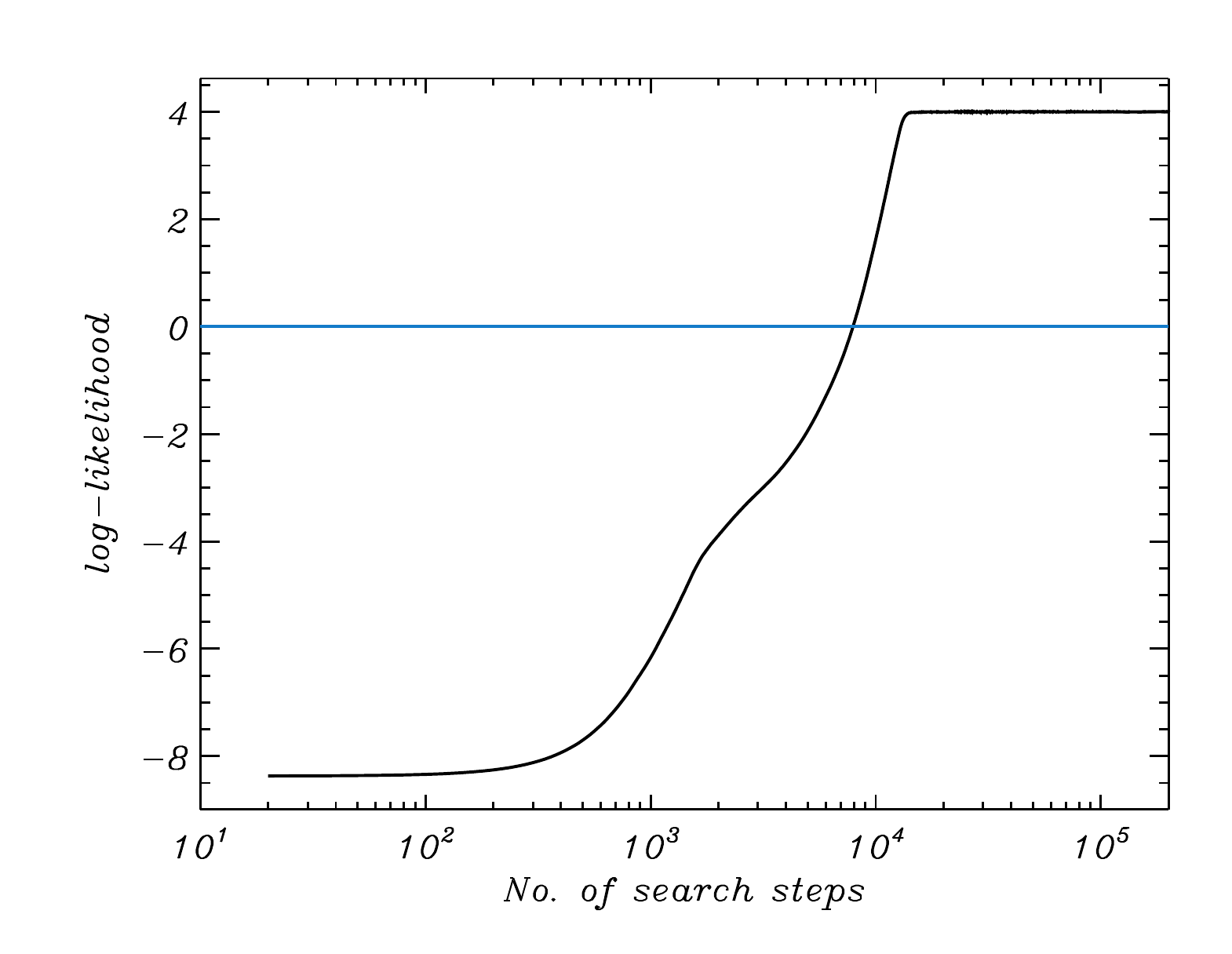}
  \includegraphics[width=0.49\textwidth]{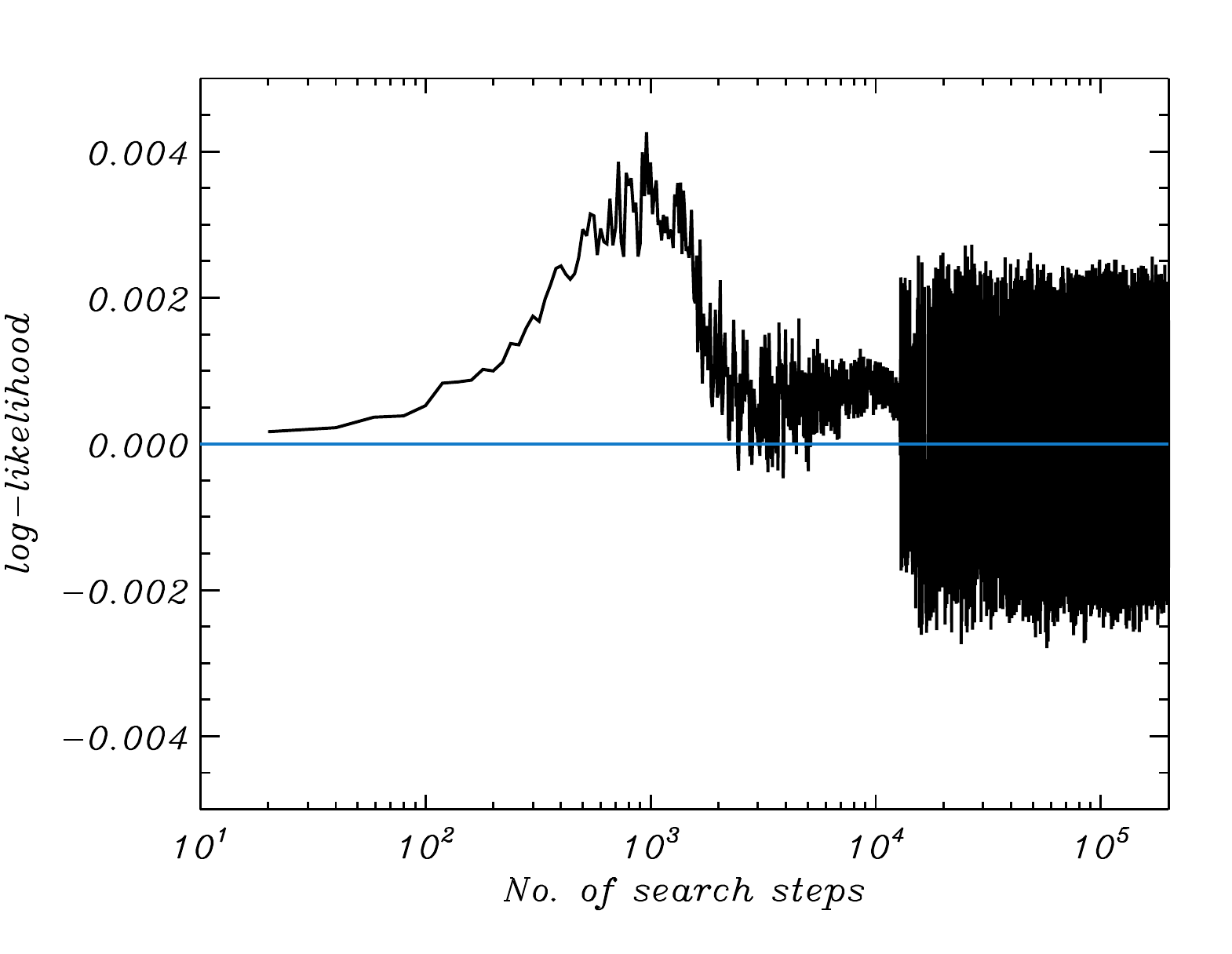}
  \caption{Left panel: the log-likelihood as a function of the iterative steps.
  Right panel: the oscillation of the log-likelihood function, presented as
  $\log(L_{i+1})-\log(L_i)$ (positive values stand for an increase in the
  likelihood.)}
  \label{fig:oscillation}
\end{figure*}

Due to chance correlations between the residual and the real common signal,
the result that gives the maximum likelihood is seldom precisely equal to the
real common signal. When the iterative steps bring us sufficiently close to the
true solution (or a local maximum), there will be a tendency for the
estimations to oscillate. This is the case in our work. Since we start from a
random white noise initial guess, the log-likelihood function will increase
monotonically at the initial stages of the search, as shown in the left panel
of figure~\ref{fig:oscillation}. However, when the RMS difference between the
``current'' estimation and the real common signal is comparable to the
residual noise, the likelihood function will start to oscillate, as shown by
the right panel of figure~\ref{fig:oscillation}.

\begin{figure*}[!htbp]
  \centering
  \includegraphics[width=0.49\textwidth]{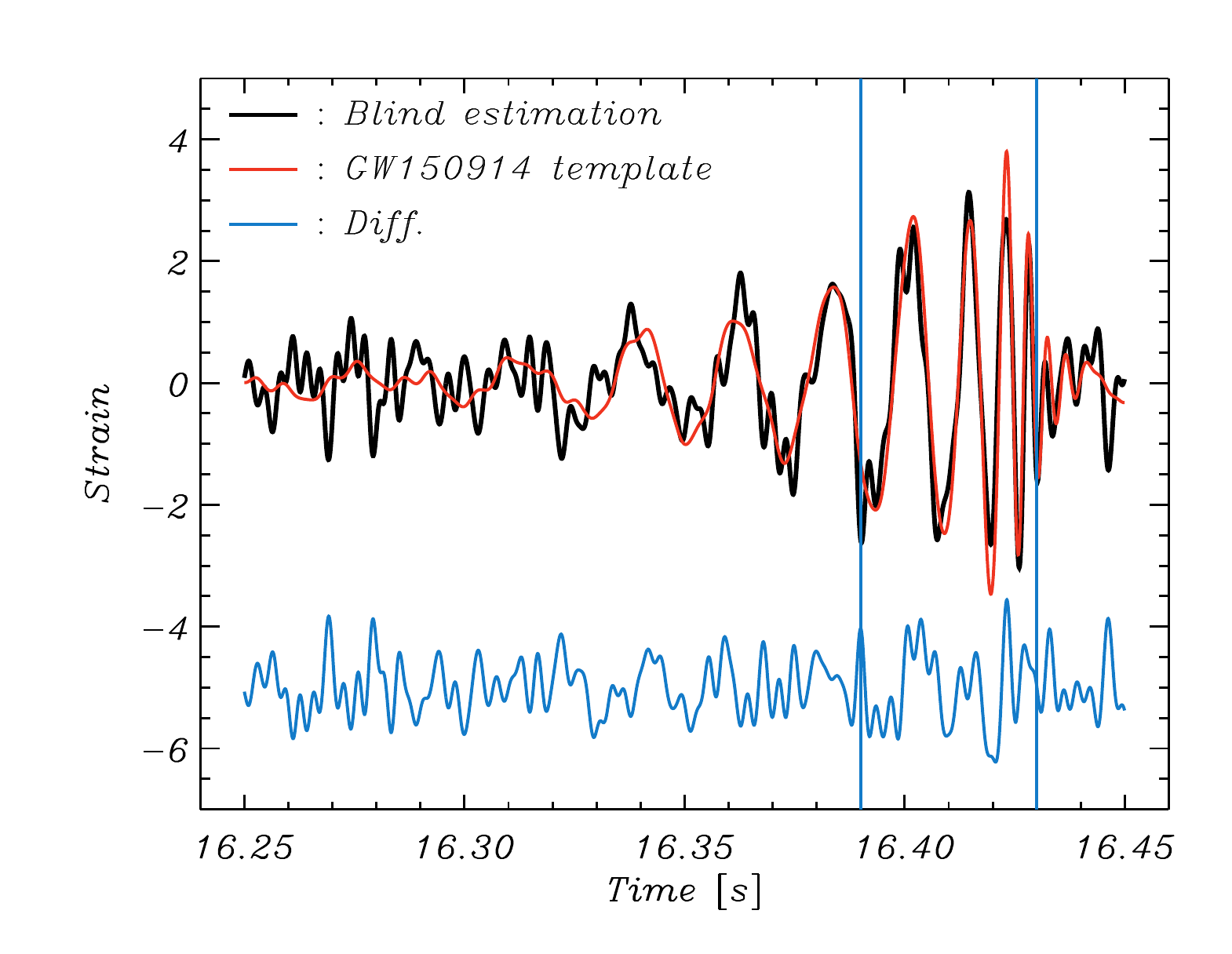}
  \includegraphics[width=0.49\textwidth]{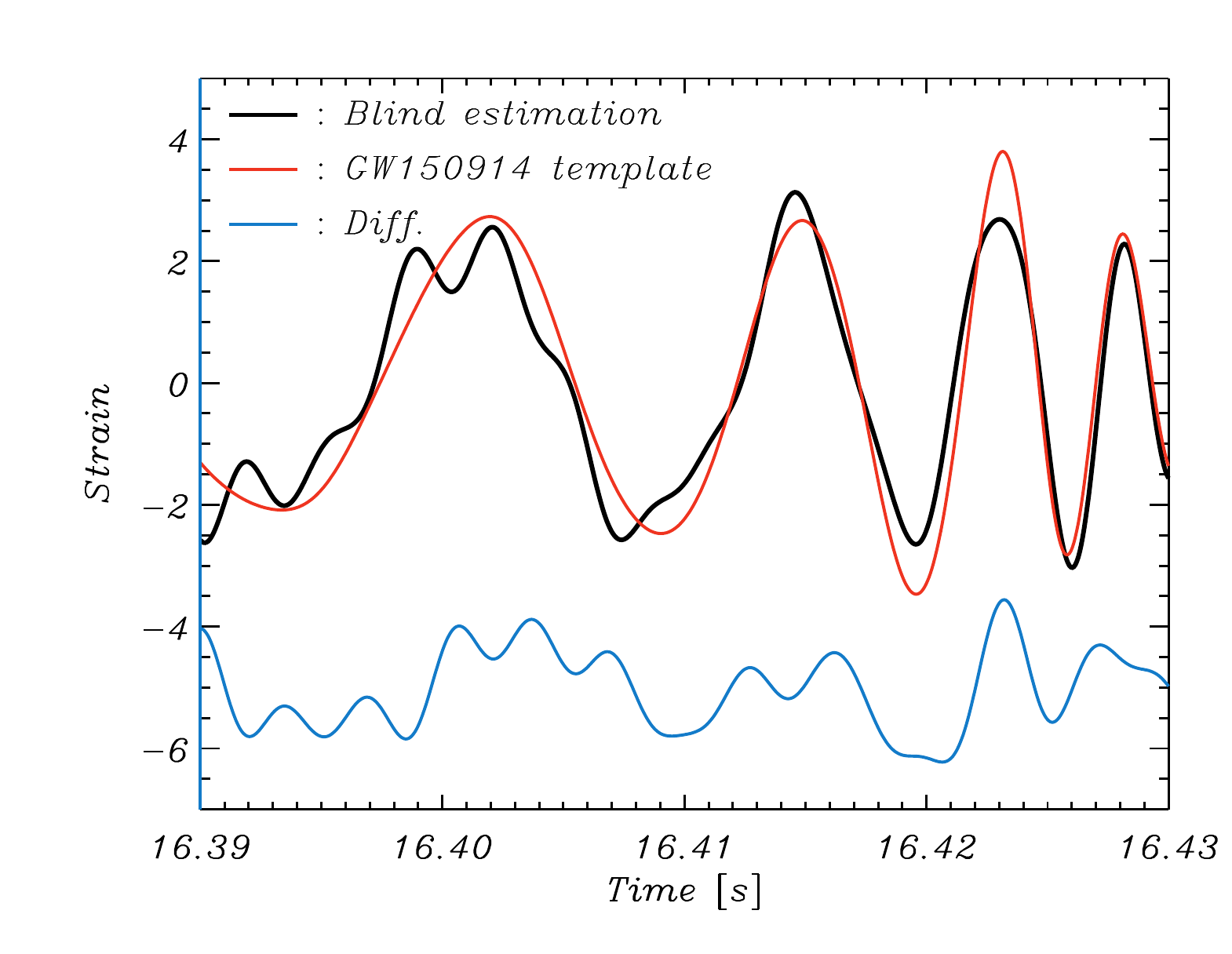}
  \caption{Our blind estimation (black) compared with the GW150914 template
  (red). Left panel: the whole 0.2~s range. Right panel: only the 0.39--0.43~s
  range (marked by the vertical lines in the left panel). The blind estimation
  denotes the average over all solutions in the second  half of a single run
  ($2\times10^5$ steps), as described in the text.} 
  \label{fig:single solution}
\end{figure*}

\begin{figure}[!htbp]
  \centering
  \includegraphics[width=0.49\textwidth]{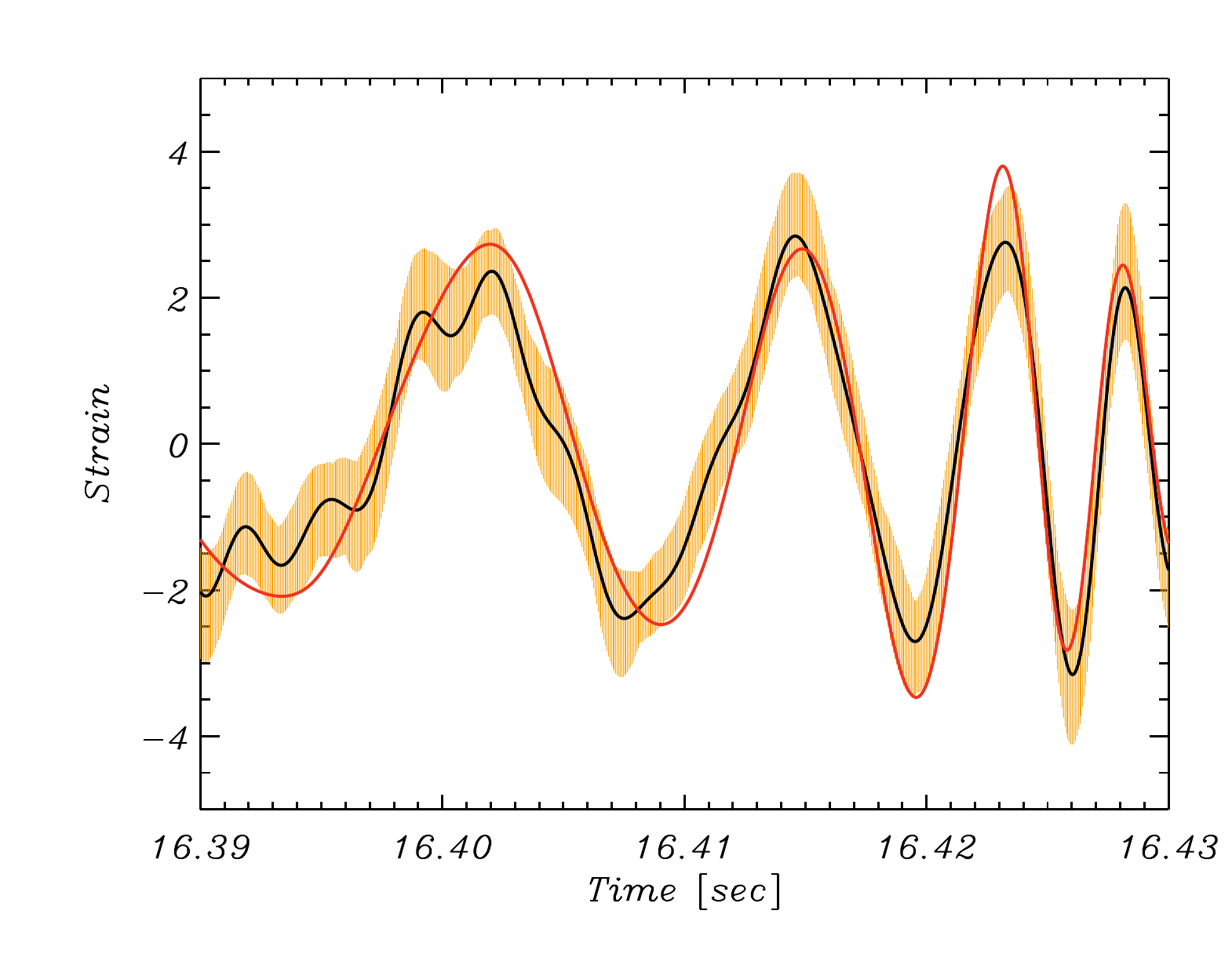}
  \caption{The minimal to maximal fluctuation
  range obtained for 100 runs (yellow band) and their average (black). Also
  shown is the GW150914 template for comparison (red). Note that the yellow
  range shown here is mainly for illustration, as described in the
  text.}\label{fig:100-runs}
\end{figure}

Therefore, in order to get a reasonable estimation of the common signal we
choose the later half of the iterative steps ($10^5$ steps, entirely in the
oscillation region). Their average is the final estimation of one
run\footnote{One run means running the blind search once, with one initial
guess, and many iterative steps following the initial guess.} as shown in
figure~\ref{fig:single solution}. Potentially, the fluctuation range for each
point can also be given by the oscillation region; however, for two reasons,
we prefer not to do this:
\begin{enumerate}
  \item As mentioned above, the number of DOF in the frequency domain is
      much smaller than the number of points in the time domain. Thus, a
      point-to-point fluctuation range in the time domain does not reflect the
      actual fluctuation range which is more tightly constrained.
  \item The fluctuation range for a single run is also related to the size
      of the iterative step, and its meaning is only relative.
\end{enumerate}
For these reasons, we choose to use only the average solution in the
oscillation region for a single run. To give an \emph{illustrative} range of
fluctuations, we repeat the procedure 100 times, starting each time with a
different random initial guess. This will yield a range of fluctuations due to
the existence of multiple local maxima, see figure~\ref{fig:100-runs}. In this
plot, we can see that the LIGO GW150914 template is not confined to the
minimal-maximal uncertainty range, indicating that this template may not
provide a reliable estimation of the common signal between Hanford and
Livingston. However, we emphasize that figure~\ref{fig:100-runs} is only an
illustration; an improved measure of the quality of the GW150914 template as
an estimator of the true common signal will be given in
section~\ref{sec:Is_GW_good_for_common_signal}.

The yellow band shown in figure~\ref{fig:100-runs} is important for two
reasons. First, it is clear that this band provides a useful measure of the
uncertainties associated with our determination of the best common signal and
that this measure could be improved by increasing the number of trials.
Second, the width of the yellow band would increase if the signal-to-noise
ratio were smaller, and it would ultimately tell us that the signal is too
weak to permit the blind extraction of a meaningful common signal.  The
existence of such an automatic warning of unreliability is an important
feature of the present method not shared by template-based analyses.

\section{Test by simulation}\label{sec:test by simulation}

We have suggested a general method for obtaining the best common signal,
applied it to the data for GW150914, and offered a rough impression of the
associated uncertainties. We will now perform a simple test of our blind
estimation method by using simulated data, where we have injected a known
common signal into a noise background, as follows:
\begin{eqnarray}\label{equ:rand_phase1}
H'(t)&& = N_H(t) + A^*(t) \\ \nonumber
L'(t)&& = N_L(t) + A^*(t),
\end{eqnarray}
where $A^*(t)$ is the known common signal, and $N_H(t)$, $N_L(t)$ are noise
backgrounds. The estimate of the common signal $A(t)\approx A^*(t)$ is
obtained by the same algorithm as presented in section~\ref{sub:search
process}.

\begin{figure*}[!htbp]
  \centering
  \includegraphics[width=0.49\textwidth]{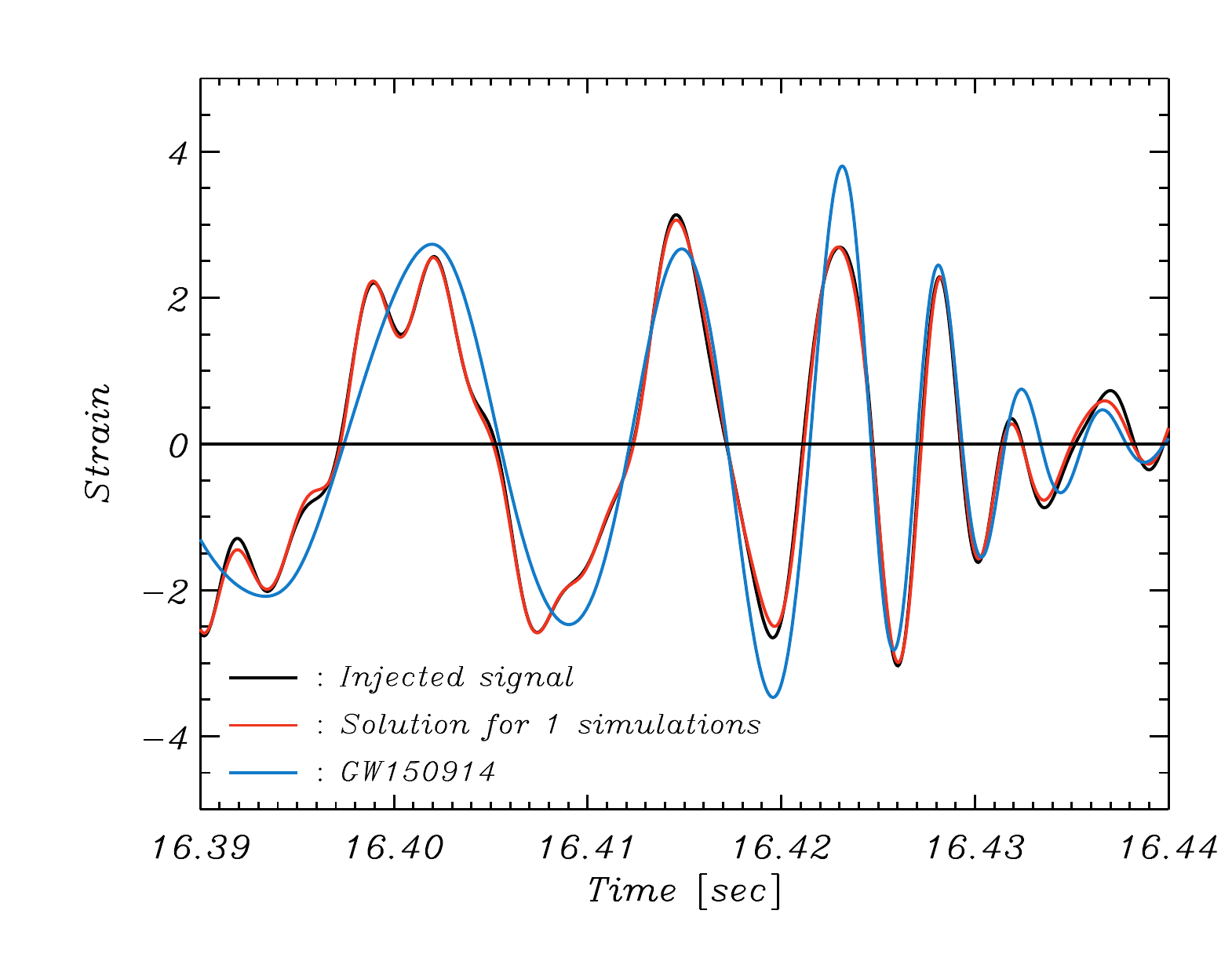}
  \includegraphics[width=0.49\textwidth]{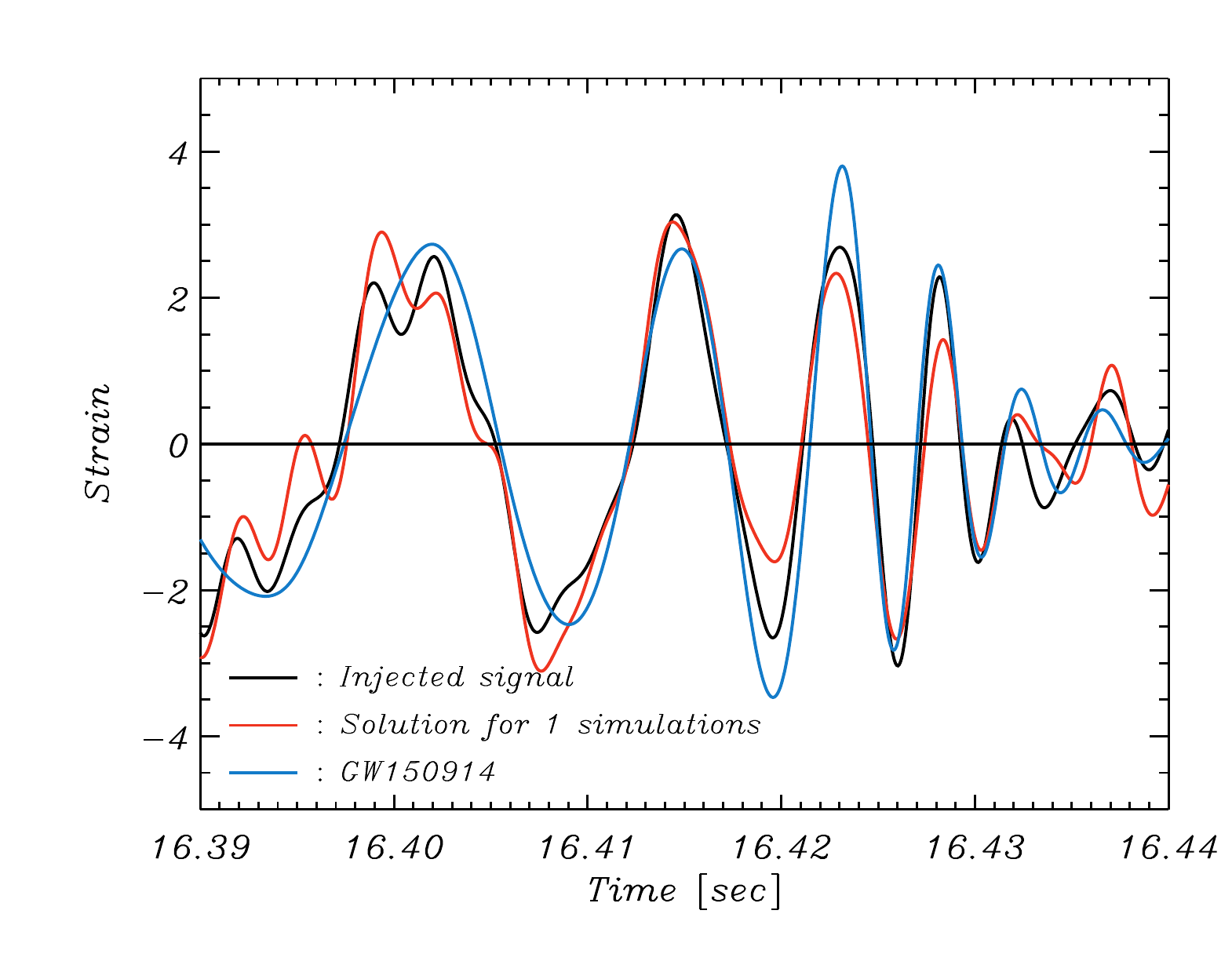}

  \includegraphics[width=0.49\textwidth]{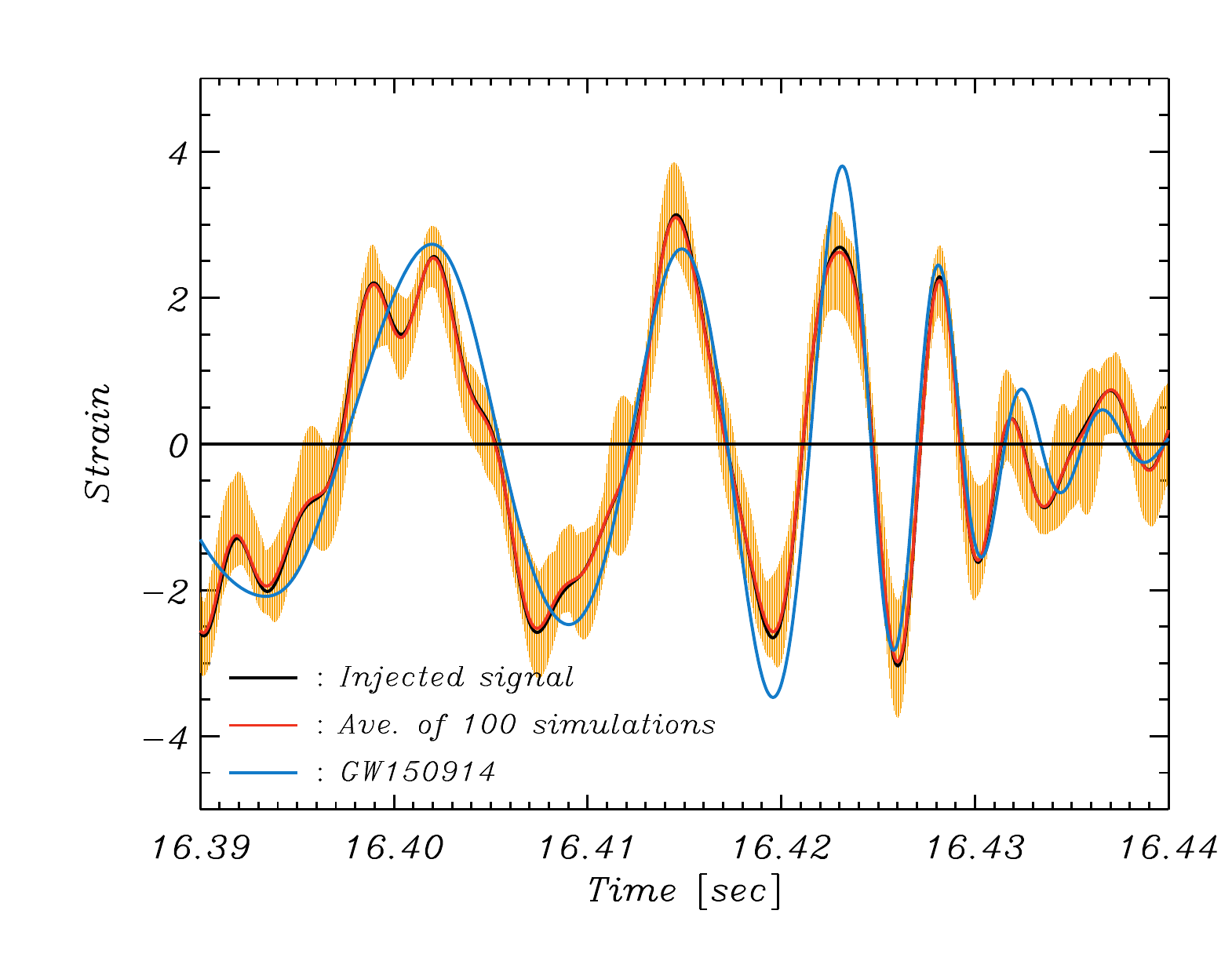}
  \includegraphics[width=0.49\textwidth]{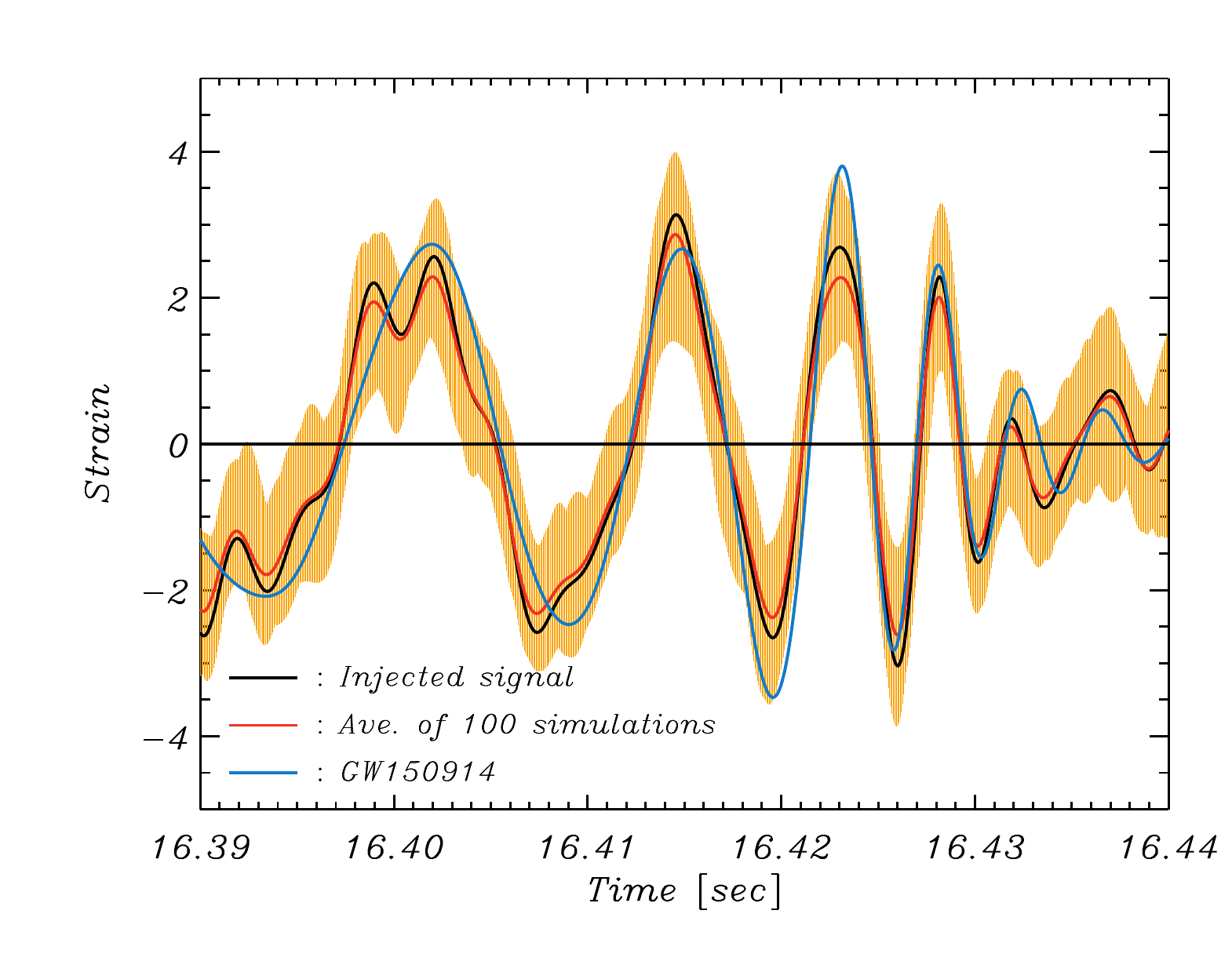}

  \includegraphics[width=0.49\textwidth]{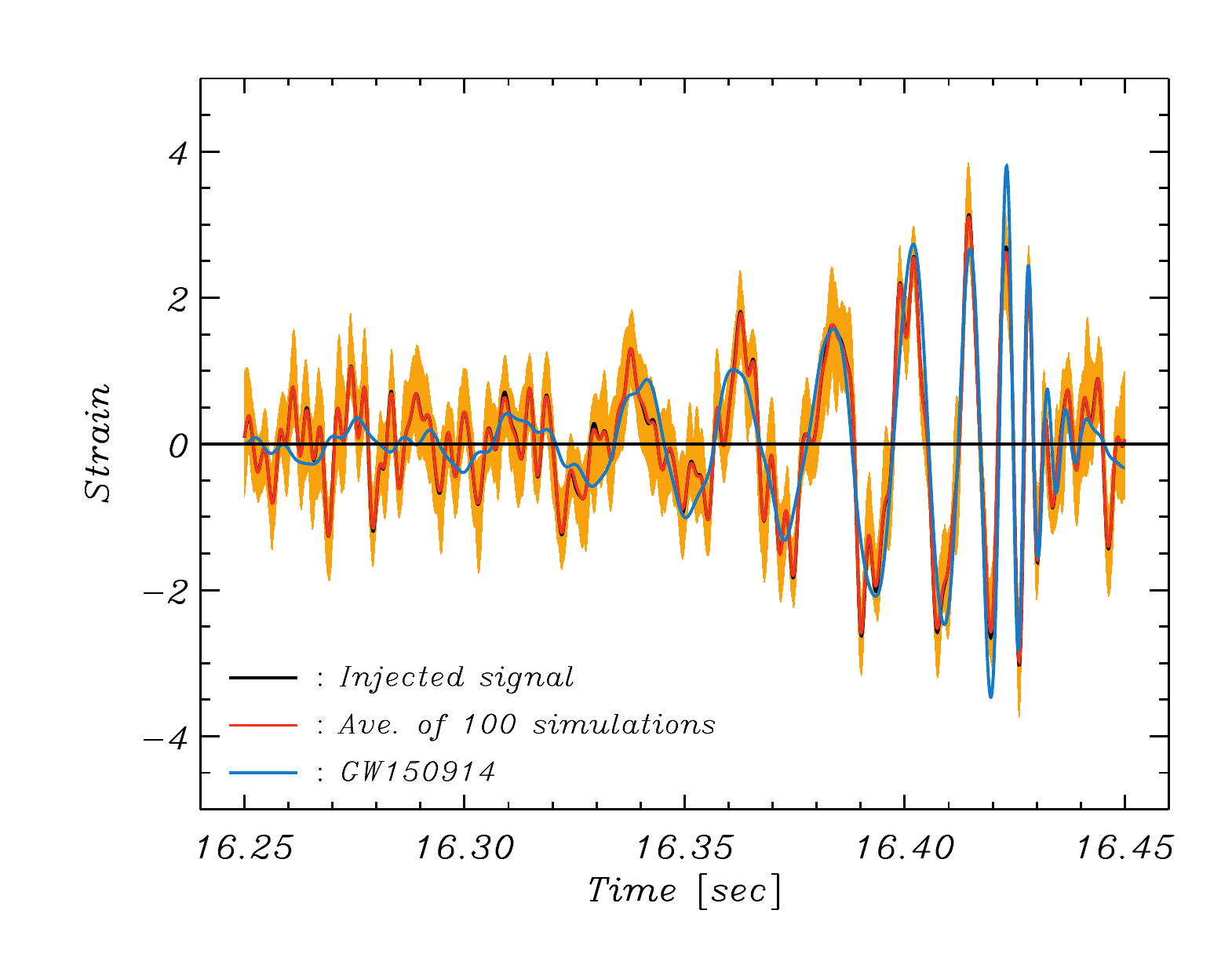}
  \includegraphics[width=0.49\textwidth]{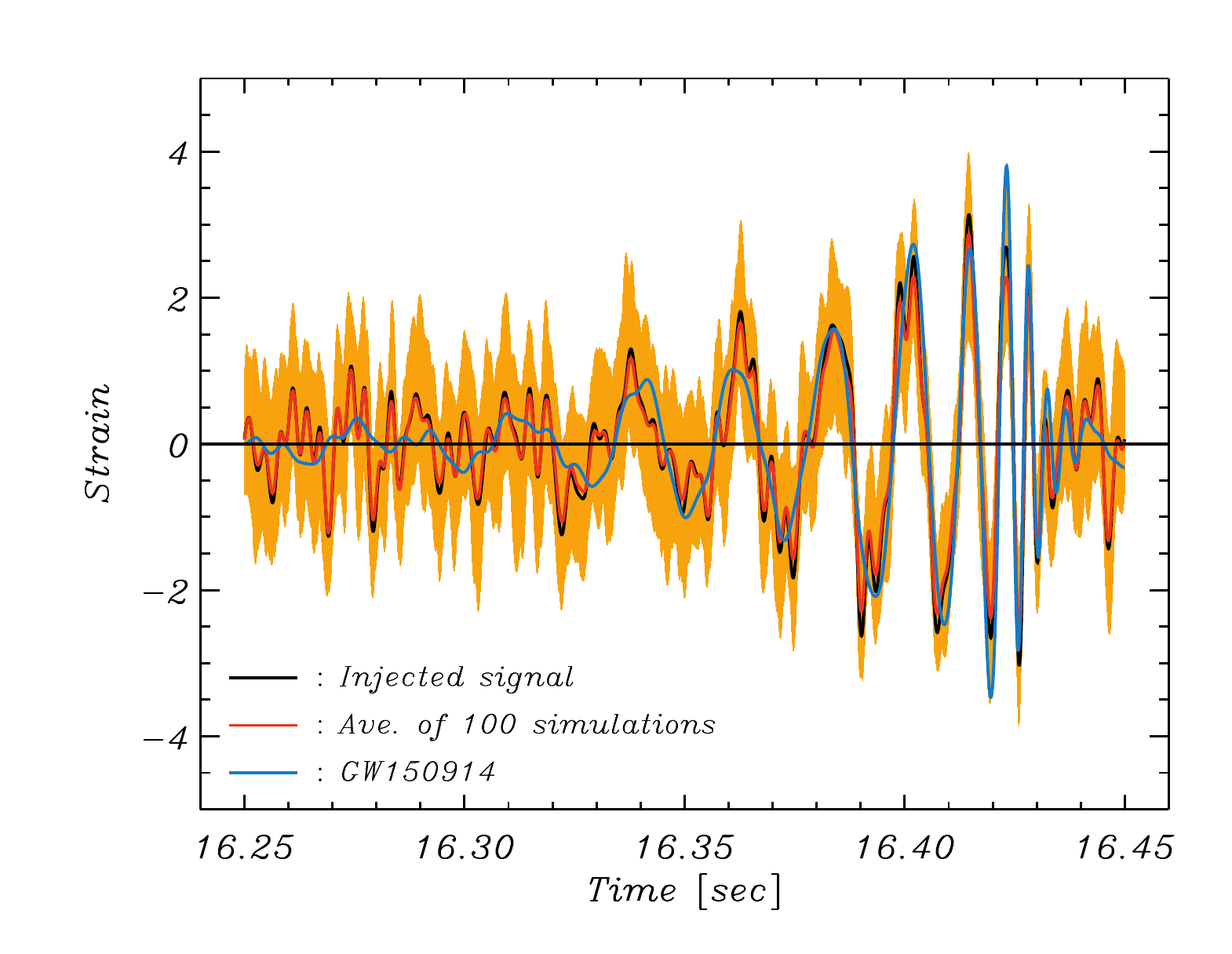}
  \caption{Results from test runs using simulated data containing the
  blind estimation from figure~\ref{fig:single solution} and real detector
  noise. In the left panels the initial guess is the injected signal itself,
  while in the right panels the initial guess is random. Top panels: One
  example simulation. Middle panels: 100 simulations, including the
  corresponding minimal to maximal fluctuation range. Bottom panels: Same as
  middle but showing the full 0.2~s range.}
  \label{fig:simulation BE-RealNoise}
\end{figure*}

Trials indicate that the method extracts the injected signal equally well,
regardless of the precise shape of the injected signal (GW150914 template or
the common signal extracted above), as well as of the noise model chosen
(random white noise or real detector noise). The largest effect --- although
still marginal --- is obtained when varying the initial guess. We therefore
present results from injecting the common signal obtained above (and shown
in figure~\ref{fig:single solution}) into real detector noise taken from
detector data at least 3 seconds away from the GW-event. We run the search 100
times, and a different segment of detector noise is chosen for each run. The
initial guess is varied, first setting it equal to the input signal, $A^*$,
and, for comparison, equal to random white noise.\footnote{In addition, new
white noise is generated for each run with a random initial guess.}
figure~\ref{fig:simulation BE-RealNoise} shows that virtually identical
solutions are obtained either way. The fact that the solution always appears
to be consistent with the known injected signal suggests that our method
provides an unbiased estimate of the best common signal.\footnote{Finally, it
is interesting that we find a smaller fluctuation range in
figure~\ref{fig:simulation BE-RealNoise} (with real detector noise from
outside the range of GW150914) than in figure~\ref{fig:100-runs} (with real
detector noise from the time of GW150914), which suggests the residual noise
associated with GW150914 is unusual.}

\begin{figure}[!htbp]
  \centering
  \includegraphics[width=0.49\textwidth]{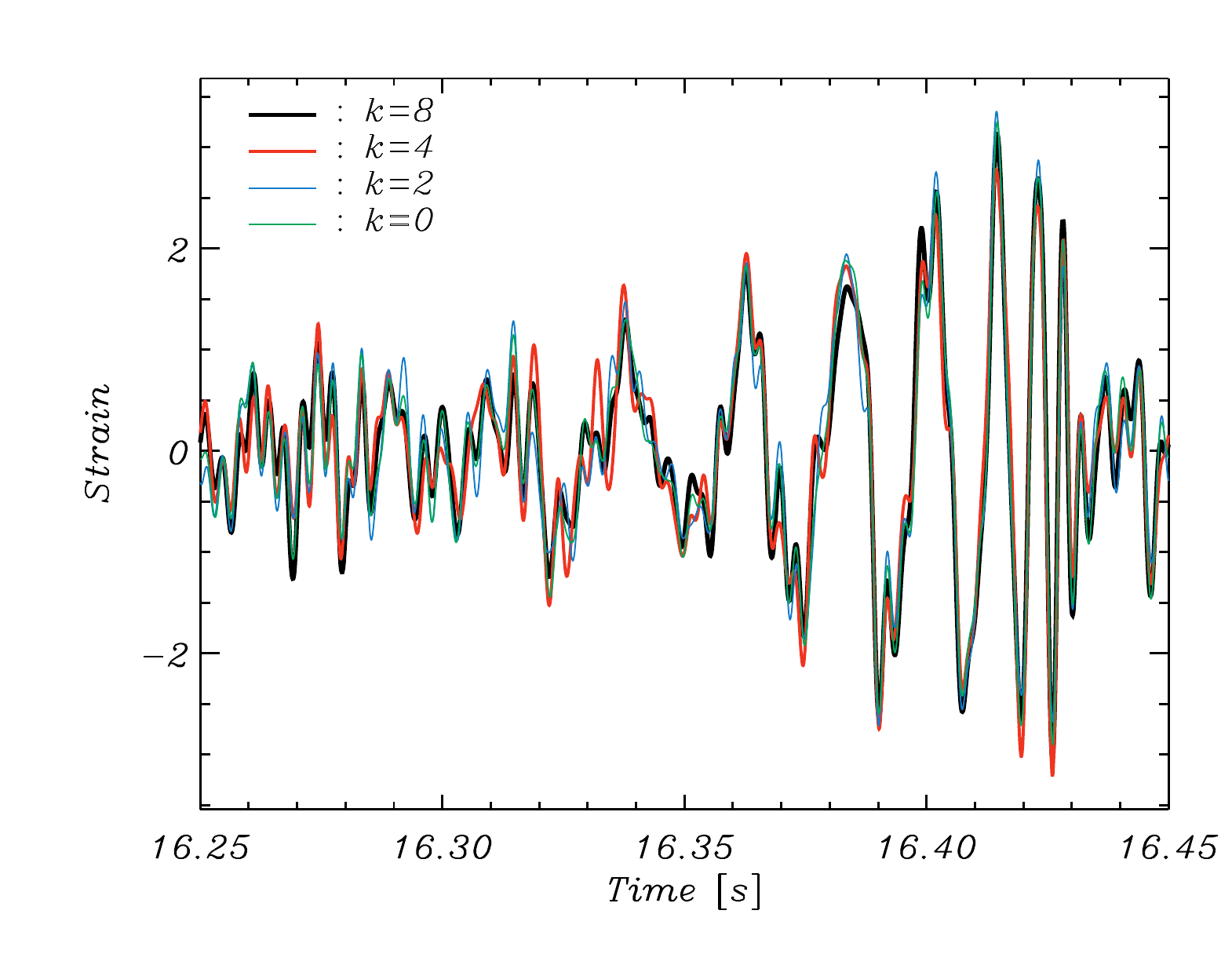}
  \includegraphics[width=0.49\textwidth]{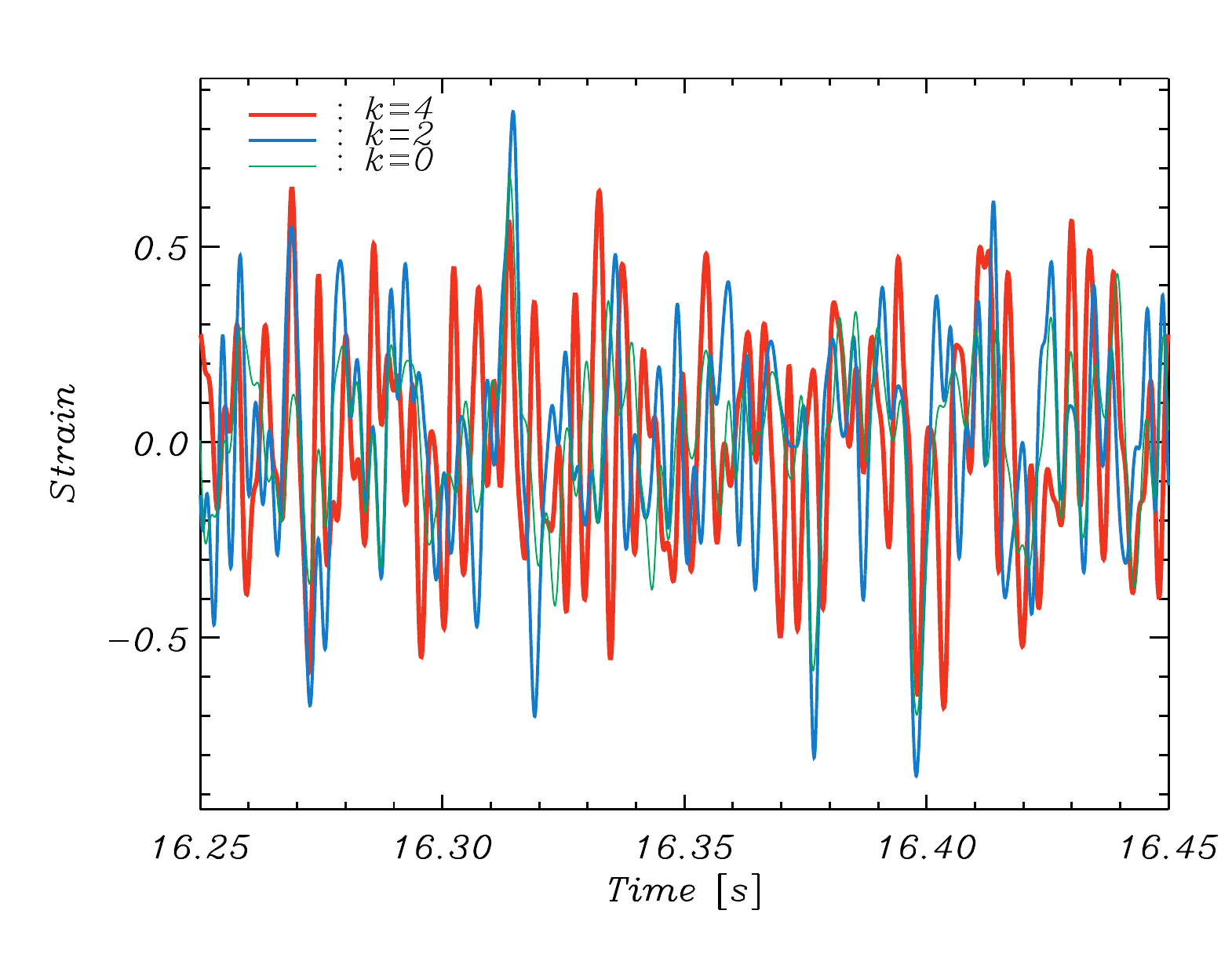}
  \caption{Test of the effect of parameter $k$ in
  eq.~\ref{equ:likelihood_simp}, in which we set $k=0,2,4,8$ respectively. The
  right panel shows the difference between these curves after subtracting the
  curve for $k=8$. Note that to concentrate on the effect of $k$, we use the
  same random initial guess for all cases. } \label{fig:effect of k}
\end{figure}

We have also performed an additional test of the influence of the parameter
$k$ in eq.~\ref{equ:likelihood_simp} in which we set $k=0,2,4,8$. The
resulting solutions are shown in figure~\ref{fig:effect of k}, from which we
can see that the extracted signals are not very sensitive to the value of $k$.
However, small differences in the template
can lead to significant differences in the residual, which can be seen in the
right panel of figure~\ref{fig:effect of k}.

\section{Testing the residuals}\label{sec:resi-test}

\subsection{The residual correlation}\label{sub:resi-cc}
\begin{figure}[!htbp]
  \centering
  \centerline{\includegraphics[width=0.49\textwidth]{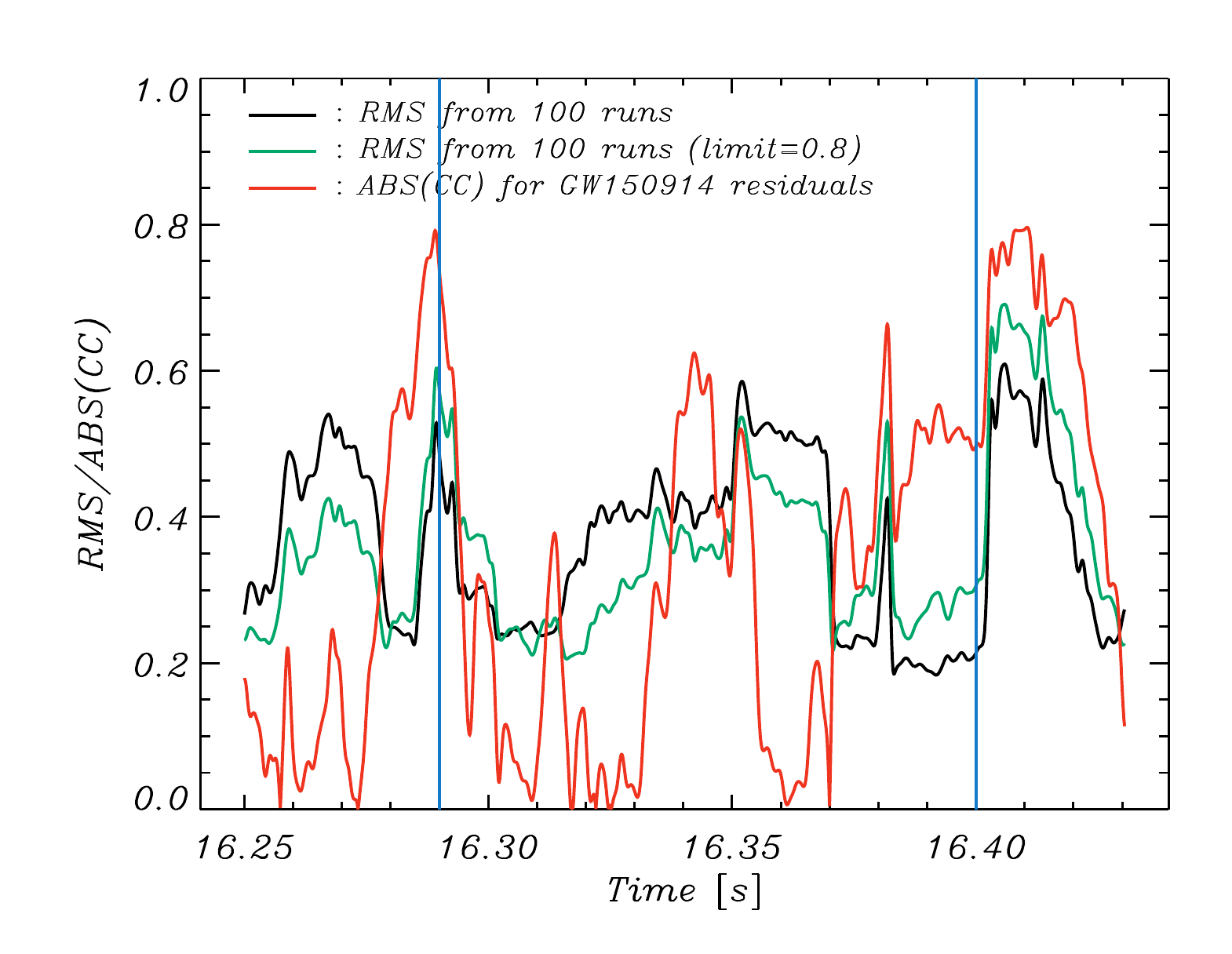}
  \includegraphics[width=0.49\textwidth]{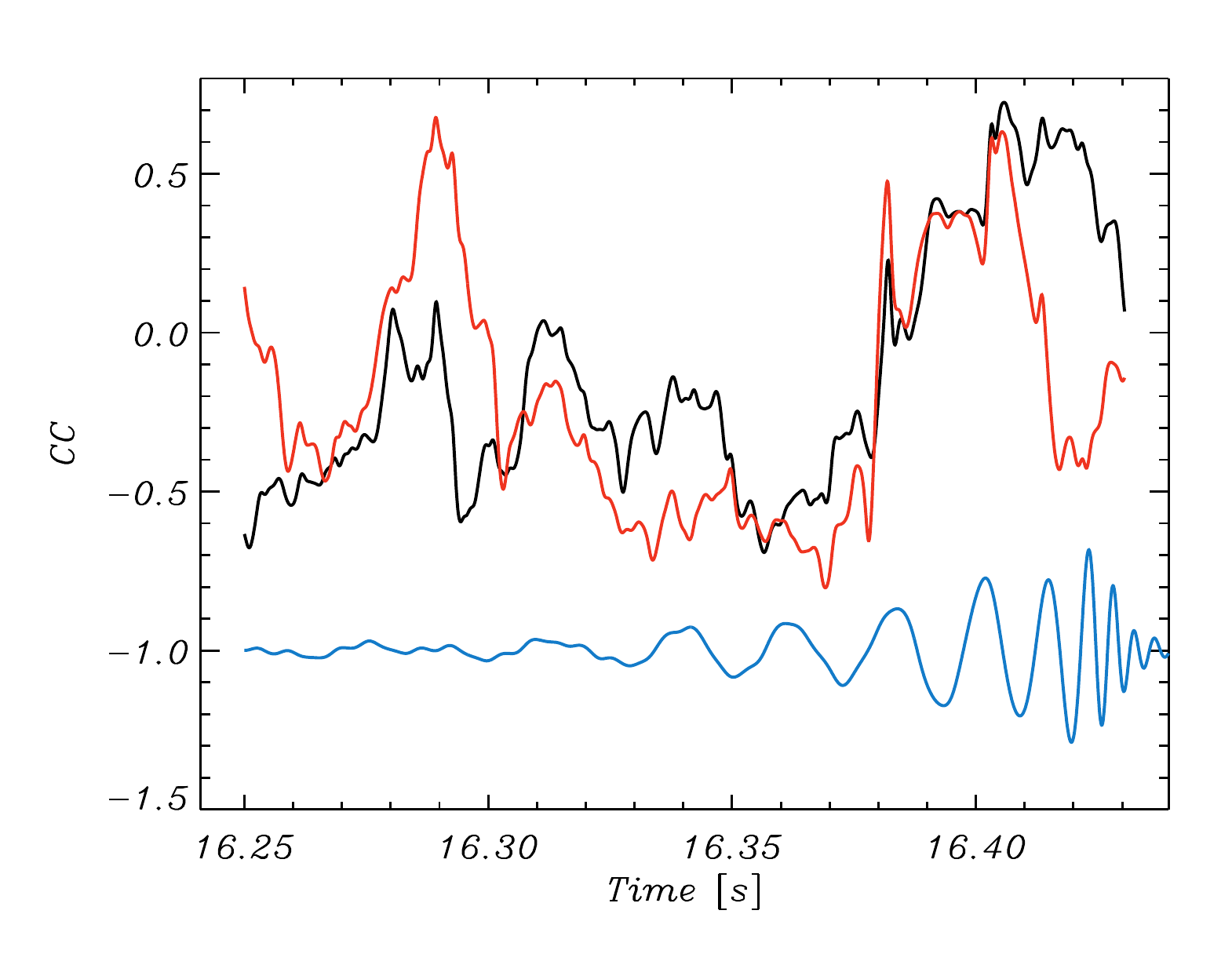}}
  \caption{The residual correlation around GW150914 after subtraction of the
  best-fit common signals and the GW150914 template. Left panel: RMS
  amplitude/absolute value of correlations, including an extra test by
  changing the threshold from 0.9 to 0.8. Right panel: Two examples of
  residual correlations from the 100 simulations calculated for a threshold of
  0.9. In addition the GW150914 template (blue) is shown for guidance.}
  \label{fig:cc-100-runs}
\end{figure}

One of the defining properties of this method is the suppression of
correlations in the residuals. After subtraction of the best common signal, we
find a cross-correlation between the residuals in Hanford and Livingston of
$0.13$ in the 0.2~s time window. For comparison, the cross-correlation in the
residuals after subtraction of the GW150914 template is $0.15$. This value
describes the correlation for the full region of investigation (in this case
0.2~s long). It is also relevant to consider this correlation on shorter time
scales (see \cite{2017JCAP...08..013C}). We can check the size of the
correlations of residuals by calculating a running window correlation between
the Hanford and Livingston residuals in each of the 100 runs (Livingston
already matched, as explained in section~\ref{sub:pre-match}):
\begin{equation}\label{equ:run win cc}
C_i(t) = \textrm{Corr} (R_{H,t}^{\hspace{10pt}
t+w},R_{\tilde{L},t}^{\hspace{8pt} t+w}),\qquad i=1,...100
\end{equation}
where $X_{t_b}^{t_e}$ denotes a data stream $X(t)$, within the time range
$t_b\leq t\leq t_e$ and we pick $w=0.02$~s to be the length of the running
window. For the following comparison we only consider the RMS amplitude of all
correlations,~$C_i$:
\begin{equation}\label{equ:mean cc amp}
\text{RMS}(t) = \sqrt{\frac{1}{N}\sum_{i=1}^N C^2_i(t)},
\end{equation}
and plot the resulting $\text{RMS}(t)$ in figure~\ref{fig:cc-100-runs} along
with the absolute value of the correlation of the residuals obtained by
subtracting LIGO's GW150914 template. We also show the average amplitude of
the cross-correlations for comparison.

We see that our common signal estimation partially reduces the residual
correlation at the precursor and main chirp regions (blue vertical lines)
though the variation of the correlation still follows the red line. This might
be explained by the existence of more than one ``common'' component, each with
a potentially different amplitude ratio between the detectors, such that they
cannot all be captured by a single estimation as here. This is similar to the
situation in CMB science, where several Galactic foreground components such as
synchrotron and thermal dust emission, each with a different spectrum, disturb
our view on the background. It is not possible to remove these effects by
assuming a single common foreground spectrum. With such an explanation, one
might conclude that GW150914 is not the only common signal between the Hanford
and Livingston detectors. The correlation for the residuals found above should
be compared with the expected correlation in 0.2~s obtained from detector
noise, which has a mean value of zero and has a standard deviation of 0.11.
The consistency of this result with the observed value of 0.13 suggests that
there is no compelling evidence for the existence of multiple common
components. Even though this possibility would merit further investigation, it
will not be considered here.

\subsection{Determination of the threshold}\label{sub:for the threshold}
The threshold used to constrain the likelihood (section~\ref{sub:likelihood})
appears to be a natural choice considering the correlation coefficient between
the GW150914 template and the Hanford data (about 0.87). We now justify this
choice by evaluating the corresponding amplitude of the residuals. Increasing
the threshold leads to solutions approaching the unwanted solution shown in
eq.~\ref{equ:unwanted solution}, which in turn leads to a decrease in the
residuals at one side (either Hanford or Livingston), which can be assessed by
computing $\sigma_{\rm{comb}}=\sigma_{R_H}\sigma_{R_L}$, where $\sigma_X$ is
the standard deviation of quantity $X$. For each of the thresholds, 0.8, 0.9,
and 0.99, we perform 100 search runs for the GW150914 data with different
random initial guesses, and find $\sigma_{\rm{comb}}$ to be
$1.44\times10^{-22}$, $1.35\times10^{-22}$, and $0.80\times10^{-22}$
respectively. For comparison, the same computations on pure noise segments of
the same length after GW150914 result in typical values between
$1.3\times10^{-22}$ and $1.6\times10^{-22}$. Thus, it is apparent that
thresholds of about 0.8 to 0.9 provide a reasonable range, whereas the 0.99 is
too high. The left panel fo figure~\ref{fig:cc-100-runs} therefore includes
results from both thresholds, 0.8 and 0.9.

We also test the average of the RMS values for thresholds 0.8 and 0.9 shown in
the left panel of figure~\ref{fig:cc-100-runs} as the green and black line.
These amount to 0.36, and 0.37, respectively. This small difference gives
further support that our results are not significantly affected by the choice
of the threshold.

\section{Goodness of the GW template as an estimator of the common
signal}\label{sec:Is_GW_good_for_common_signal} 

In this section, we wish to provide a quantitative evaluation of the quality
of the GW150914 template as an estimator of the common signal between Hanford
and Livingston. This evaluation will be based on the likelihood of the
template calculated by eq.~\ref{equ:likelihood_simp}.

First, we point out that the likelihood for the GW150914 template is 3.07 and
the likelihood for the blind estimation shown in figure~\ref{fig:single
solution} is 3.99. While this difference already suggests that the quality of
the GW150914 template is not high, these two numbers do not tell us the
probability for accepting the template. Thus, we use the following two
approaches to address this question.

The first approach is to take the GW150914 template as the initial guess to
start the search. If the difference between 3.07 and 3.99 is insignificant, we
should see oscillations from the beginning. However, figure~\ref{fig:IG is
GW150914} shows a clear monotonic rise in the likelihood and a range of final
oscillations that is much less that the difference between 3.99 and 3.07. This
provides a qualitative indication that the difference is significant and
that GW150914 is not a good estimator of the common signal.

\begin{figure}[!htbp]
  \centering
  \includegraphics[width=0.49\textwidth]{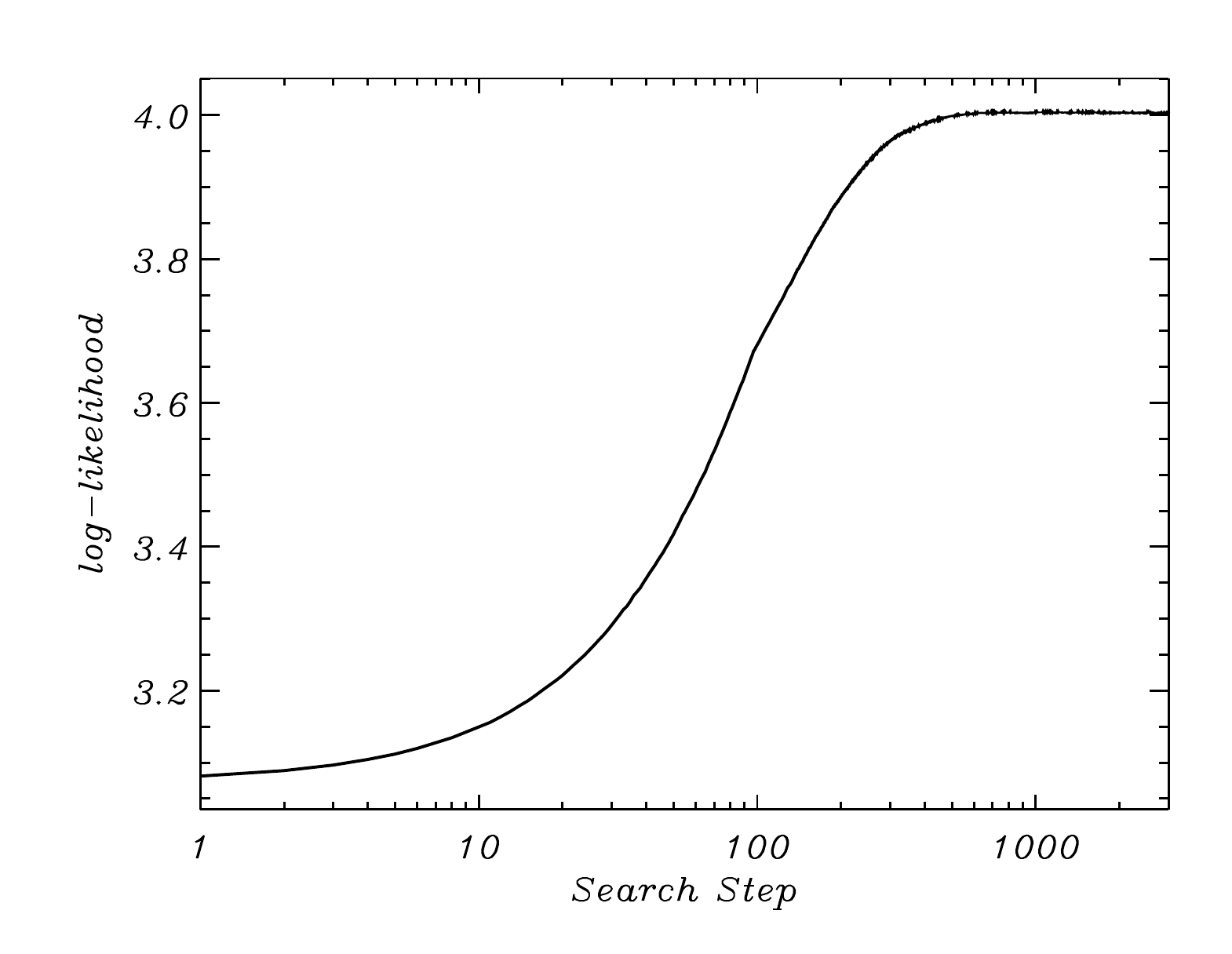}
  \caption{Evolution of the likelihood when selecting the GW150914 template as
  the initial guess.} \label{fig:IG is GW150914}
\end{figure}

The second approach is to run the search with steps, $d(t)$, that are
sufficiently large to enable the solution to jump from one local maximum to
another and thus prevent convergence to a single local maximum. We can then
find the probability of obtaining a likelihood of 3.07 or less. This will be
related to the probability that the GW150914 template should be rejected as
the best common signal.

The key issue of this second approach is to ensure that the iterative steps
are large enough to be able to jump between local maxima. This can be
monitored by checking the likelihood of the average solution: if all steps are
oscillating around only one local minimum, then the average solution will also
have a high likelihood, like the one in figure~\ref{fig:single solution}. On
the other hand, if the average solution has a very low likelihood, then the
iterative steps must be jumping between multiple local maxima. Working in
Fourier space, we empirically were led to adopt a step size roughly $10\%$ of
the expectation for the Fourier amplitude, and $1\%$ of the whole size of the
parameter space for the Fourier phase.

\begin{figure*}[!htbp]
  \centering
  \includegraphics[width=0.49\textwidth]{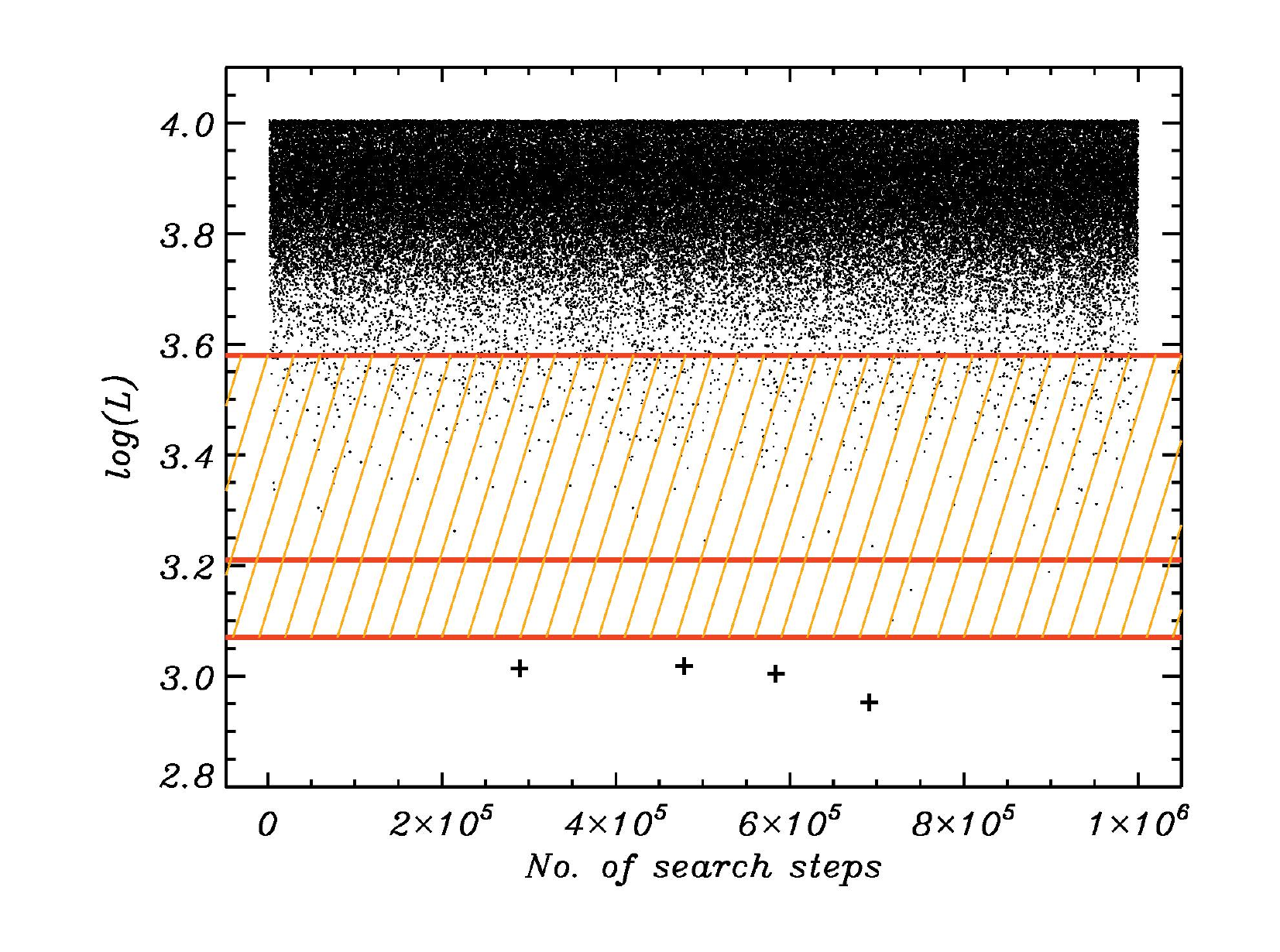}
  \includegraphics[width=0.49\textwidth]{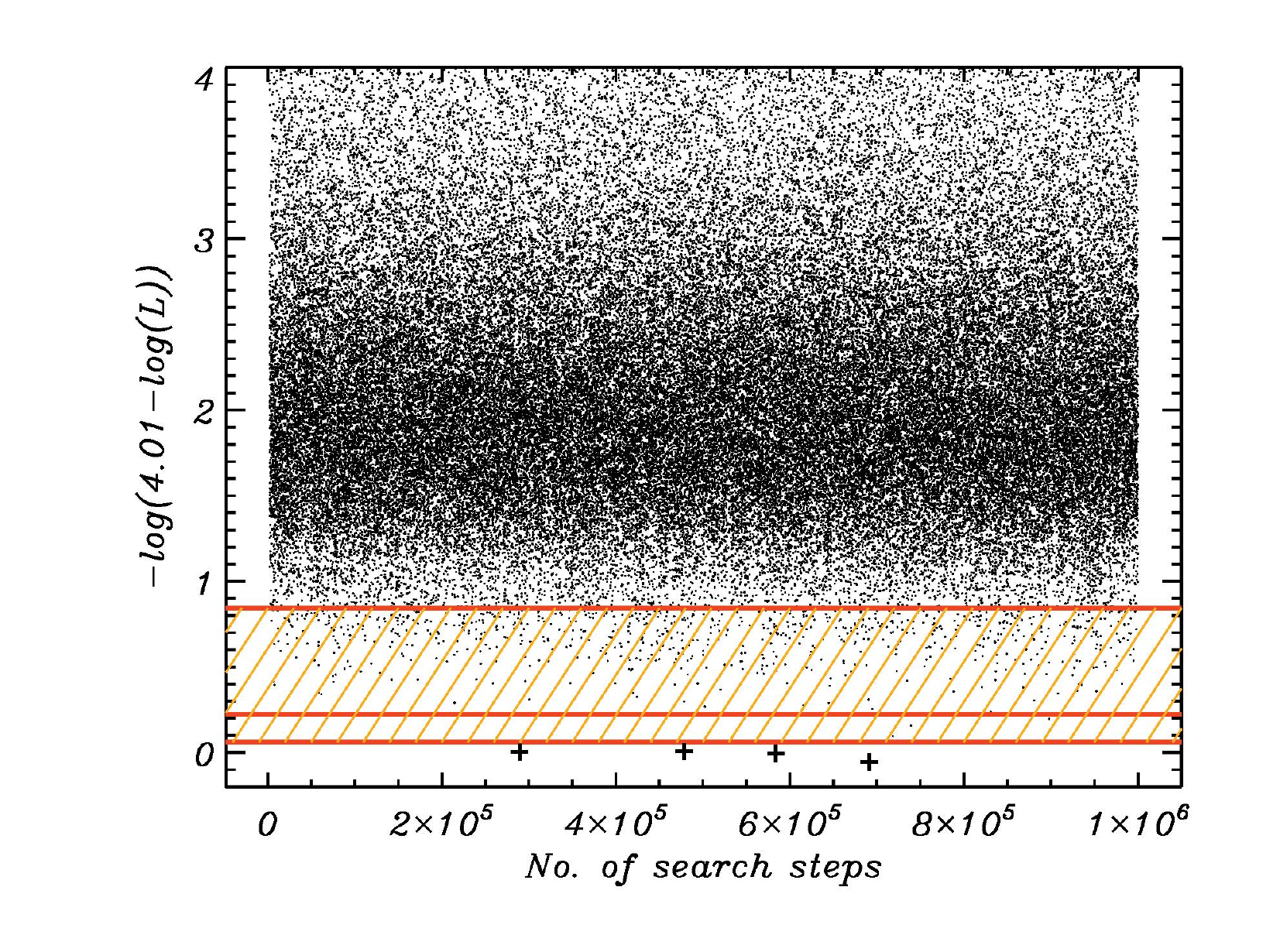}
  \caption{Likelihood as function of steps when the size of the steps are big
  enough to ensure jumping between local maxima. The horizontal line marks the
  likelihood of the GW150914 template. Both panels are identical, only that in
  the right panel we use the transformation $-\log(4.01-\log(L))$ to make the
  points close to the likelihood peak more visible. The three horizontal lines 
  from bottom to top mark the likelihood of the GW150914 template, a large total 
  mass template, and the maximum likelihood gravitational wave template, 
  respectively. (See text for description.) It is clear that the
  GW150914 template is at a position with very low likelihood. The shaded yellow 
  region indicates the range of likelihoods obtained from different gravitational 
  wave templates
  considered here.}
  \label{fig:likelihood for big steps}
\end{figure*}

\begin{figure}[!htbp]
  \centering
  \includegraphics[width=0.49\textwidth]{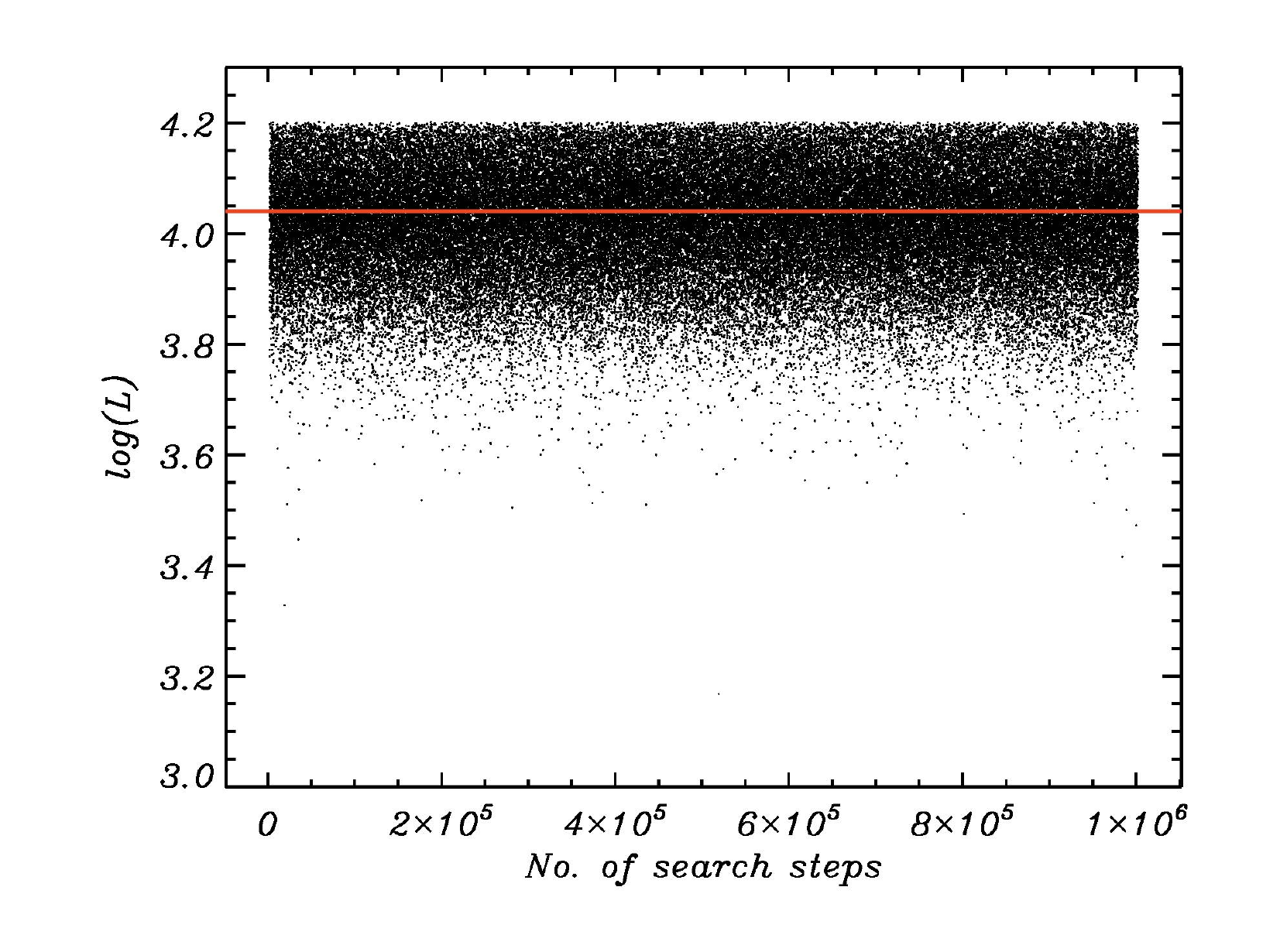}
  \caption{Same as the left panel of figure~\ref{fig:likelihood for big steps},
  but for simulation with the best common signal (see figure~\ref{fig:single
  solution})
  and real detector noise.} \label{fig:likelihood for big steps-sim}
\end{figure}

A run of $10^6$ iterative steps with this step size resulted in an average
signal of all solutions that gives a likelihood of around 0.1. This is much
lower than either 3.99 or 3.07 --- sufficiently low to ensure that the
iterative steps are jumping between multiple local maxima. All likelihoods are
plotted in figure~\ref{fig:likelihood for big steps}, from which we see that
almost all steps are above the 3.07 line (the lowest of the horizontal, red
lines). There are only 4 instances with a likelihood lower than 3.07. This
suggests that the GW150914 template is preferred only with a probability of
around $4\times10^{-6}$ compared to the best common signal found here. In
figure~\ref{fig:likelihood for big steps-sim} we also present a simulation
using a signal composed of the best common signal of figure~\ref{fig:single
solution} and real noise taken from the data outside the GW150914 domain. In
this case we see that the likelihood of the input signal (horizontal, red
line) is consistent with the results in the region of oscillations.

Despite its appearance in figure~1 of ref. \cite{PhysRevLett.116.061102}, the
GW150914 template discussed so far is not the best-fit template. We have carried
out a search to investigate the possibility that other templates are more
likely according to our method. Although the dependence of template morphology
on black hole masses and spins is highly degenerate in the vicinity of
GW150914 \cite{Degeneracy}, changes in the masses orthogonal to the chirp mass
degeneracy can still affect the results.

By nature of its definition, the log-likelihood is much more sensitive than
the cross-correlation to small changes in morphology.  This sensitivity can be
misleading since chance correlations can have a significant effect on the
log-likelihood when the probability of a given template is small. Therefore,
in our search, we have sought to maximize the log-likelihood by performing a
Monte Carlo search over several thousand choices of black hole
templates\footnote{In the following, we employ templates generated using PyCBC
with \texttt{SEOBNRv2}~\citep{2014PhRvD..90h2004D, 2016CQGra..33u5004U,
https://doi.org/10.5281/zenodo.1058970,2014PhRvD..89f1502T}. \texttt{SEOBNRv2}
was used by LIGO for analysing GW150914~\citep{TheLIGOScientific:2016wfe}. A
more recent version, \texttt{SEOBNRv4}~\citep{Bohe:2016gbl}, yields the same
conclusions.} with various masses and spins and have optimized the matching
parameters $\Delta$ and $\tau$ to maximize the log-likelihood. A general
feature of our search is that templates with component masses somewhat higher
than the quoted values of 36 and 29 \cite{PhysRevLett.116.241102}, and with
anti-aligned spins, often give higher log-likelihoods.  Specifically, the
parameters $m_1 = 40$, $m_2 = 38, \chi_1= 0.96$, and $\chi_2 = -0.85$
($m_{1/2}$ are the masses of the two black holes in units of $M_\odot$, and
$\chi_{1/2}$ are their dimensionless spins) yield a maximum log-likelihood of
3.58 (shown as the highest red line in figure~\ref{fig:likelihood for big
steps}) after appropriate bandpassing, notching, and  matching. This
corresponds to a $p$-value of 0.008, which is the estimated probability that
the best common signal is a template describing the merger of black holes.

To study further the effect of increased masses, we change the parameters of
the previous template to $m_1 = 48, m_2 = 38, \chi_1 = 0.96, \chi_2 = -0.85$,
as an example, which yields a log-likelihood of 3.21 (also shown in
figure~\ref{fig:likelihood for big steps}). We note that such a template gives
a high SNR in the Hanford and Livingston data streams, and thus appears to be
consistent with one of LIGO's methods for finding a best-fit template.

\section{Discussion}\label{sec:discussion}

The precise determination of the morphology of detected gravitational waves,
and particularly their deviation from theoretical templates, is important for
understanding the physics of binary mergers. For example, binary mergers can
be perturbed by tidal interactions with embedded stars, especially in neutron
star mergers \cite{2017arXiv171102644A}. However, signal-extraction methods
based on template banks are inherently biased and can overlook such features.
Furthermore, template-based methods, if not handled carefully, can lead to
misunderstanding or overestimation of detection significance.

In this paper we have developed a new template-free method for extraction of
the common signal in two detectors. Our method is based on a likelihood
approach that minimizes the Pearson cross-correlation between the residuals
and maximizes the cross-correlations between the estimated common signal and
the strain data. The approach is nonlinear by design and, unlike most common
linear models for combining signal and noise, it gives a family of solutions
differing as a consequence of the inevitable chance correlations between each
member of the family and the corresponding residuals. This method makes no
assumptions about the statistical properties of the noise, and cross-
correlations are calculated from the data sets alone.

We have used the published 0.2~s Hanford and Livingston strain records for the
GW150914 event \cite{2015JPhCS.610a2021V}. We have found that the morphology
of the family of common signals $A_i(t)$ is different from the morphology of
the GW150914 template. This template deviates from the maximal-minimal
uncertainty range determined by the spread in the members of the family, as
illustrated in figure~\ref{fig:simulation BE-RealNoise}. This deviation is
particularly visible in the final peaks, which are incongruously high in the
GW150914 template. Furthermore, one can see peculiarities in the common
signals in the vicinity of the nodes, where the theoretical template is zero
and the corresponding data are supposedly only noise.

In order to test the stability of our method, we have performed simulations
using noise taken from Hanford and Livingston strain data well outside the
GW150914 event together with injected known waveforms constructed in 100 runs
of the blind search method in order to compute the fluctuation range. The
comparison between injected and reconstructed waveforms, presented in
figure~\ref{fig:simulation BE-RealNoise}, shows the stability of the method.

We are especially interested in the cross-correlation between the residuals
resulting from subtraction of the common signal. Since the family of solutions
$A_i(t)$ leads to a family of residuals, we have considered an average over
100 runs for $A(t)$ to get the moving window cross-correlations presented in
figure~\ref{fig:cc-100-runs}. We find a reduction of the cross-correlations in
the chirp domain, detected in \cite{2017JCAP...08..013C}, with preservation of
the precursor and the echo effects discussed therein. This suggests that the
results in \cite{2017JCAP...08..013C} can be explained in part as a
consequence of the subtraction of the GW150914 template instead of the best
common signal found in this work.\footnote{We would also like to mention a
recent work \cite{2017arXiv171100347G} devoted to reconstruction of the common
waveform from the LIGO/Virgo data. This method is based on the assumption of
Gaussian noise, and more importantly, leads to a specific non-coherent signal
with increasing amplitudes getting approaching to the chirp time domain. The
conclusion made in~\cite{2017arXiv171100347G} referring to the residual
cross-correlations presented in~\cite{2017JCAP...08..013C} as chance
correlations in the context of stationary Gaussian noise, stands in
contradiction to our finding of a corresponding $p$-value of $4\times10^{-3}$.
Moreover, considering the non-stationarity of the detector noise, it is very
challenging even to define these chance cross-correlations.}

To compare the quality of the derived waveforms with the GW150914 template, we
have proposed a goodness of fit test based on the likelihood approach. The
idea is to use the GW150914 template as the initial guess in the search for
the best common signal. If this template is close to the best fit solution, we
should see oscillations of $A_i(t)$ around that template. However, we have
found that this is not the case. In particular, the corresponding likelihood
for the initial template is 3.07, while the value for the best fit solution
oscillates around 3.99.  Such a gap in the likelihood strongly disfavors the
GW150914 template.  In addition, we have considered a large selection of
templates based on other black hole parameters.  The largest likelihood found
was $3.58$, which is significantly smaller than that of the best common signal
found by the blind search algorithm. This result corresponds to a $p$-value of
$0.008$. In fairness, it should be noted that this probability is a purely
statistical result based on the assumption that the details of the physical
description of the black hole merger are exact.  It is possible that the
relatively small difference between LIGO's best fit template and our best
common signal is due in part to theoretical uncertainties associated with
template calculations. Nevertheless, it is our view that this result
emphasizes the essential importance of minimizing the cross-correlations
between the residuals as well as maximizing the cross-correlations between a
template and the signals from each of the two detectors.

It must be remarked, however, that the method employed requires certain
properties of the signal to be true. By construction, the extracted common
signal must be the only signal common to the two detectors. If more common
components are present in the two detectors, our method will try to
accommodate these into its best-fit common signal extraction and thereby also
distort the estimation. The resulting compromises in the common signal would
necessarily lead to correlations in the residuals larger than what would be
expected for statistically independent noise. Furthermore,
eq.~\ref{equ:template transit} does not allow for precessing or elliptical
binaries, nor does it respect higher-order modes in the gravitational waves
(see also~\citep{2017arXiv170909181H}). Additionally, there could be
statistical and systematic errors in creating the GW-data from the detector
channels~\citep{PhysRevD.95.062003}. In the event that such corrections were
necessary, the present results would change accordingly. Evidently, similar
corrections would also have to be incorporated in much of LIGO/Virgo's
analysis that depends on matched filtering or the use of simplified template
models.

Redundancy is a powerful tool for the extraction of weak signals in the
presence of noise, and it is widely recognized as a crucial element in the
detection gravitational waves. It is less obvious how this redundancy is best
exploited. Template analyses can provide estimates of physical parameters
given the certain knowledge that the true template is in the set considered.
At their worst, template analysis can lead to a misidentification of the
nature of candidate events.  In our view, the only acceptable strategy is one
that first finds the best common signal from the data alone using methods that
are free of theoretical preconceptions and then makes the comparison to the
predictions of specific physical models.  We have presented one such approach
here and used it to analyze the data for GW150914.  Cases in which the
signals are too weak to permit unbiased determination and which therefore
require the use of templates for signal detection should be regarded with
extreme caution: It is likely that the conclusions of such analyses will be
determined by theoretical preconceptions and not by the data itself.

\acknowledgments{ 

This research has made use of data, software and/or web, tools obtained from
the LIGO/Virgo Open Science Center (https://losc.ligo.org), a service of LIGO
Laboratory, the LIGO/Virgo Scientific Collaboration and the Virgo Collaboration.
This work was partially funded by the Danish National Research Foundation
(DNRF) through establishment of the Discovery Center and the Villum Fonden
through the Deep Space project. Hao Liu is supported by the Youth Innovation
Promotion Association, CAS.

}

\providecommand{\href}[2]{#2}\begingroup\raggedright\endgroup

% \bibliography{newbib}

\end{document}